\documentclass{amsart}

\usepackage{style}
\usepackage[font=footnotesize]{caption, subcaption}
\usepackage{xcolor}
\usepackage{soul}
\usepackage{todonotes}

\title[GNN based non-local models: towards RANS turbulence modeling]{Graph Neural Network based non-local models: Toward application to RANS turbulence modeling}
\author{Salvatore Civiello$^{a, b}$, Davide Oberto$^{b}$, Antonia Larese$^{a, c}$,  Gianluigi Rozza$^{b}$}

\address{$^{a}$ Department of Mathematics “Tullio Levi Civita”, University of Padova, Padova, 35121, Italy}
\address{$^{b}$ Mathlab, Mathematics area, SISSA, Via Bonomea 265, 34136, Trieste, Italy.}
\address{$^{c}$ Institute for Advanced Studies, Technical University of Munich, Garching, 85748, Germany}

\email{\{sciviell, doberto, grozza\}@sissa.it}

\begin{document}

\begin{abstract}
Most machine-learning closures for the Reynolds-averaged Navier-Stokes (RANS) equations are local, predicting the closure quantities at each point solely from local flow features, despite the non-local nature of turbulent interactions. We propose a non-local closure framework based on Graph Convolutional Networks (GCNs) that exploits the mesh connectivity. Each finite-volume cell becomes a graph node, and cells sharing a face are connected by an edge. Through message passing, each layer aggregates information from neighboring nodes, so that after $k$ layers the prediction at a cell depends on its $k$-hop neighborhood, which naturally accounts for non-local effects. To make training feasible when each simulation yields a single large graph, we also introduce a breadth-first-search $k$-hop subgraph sampling strategy, which acts as data augmentation and cuts the cost per processed graph during training. We compare GCNs with feed-forward networks (FNNs) of equal size on two periodic-hills benchmarks: a laminar advection-diffusion transport problem that mimics the non-local character of RANS closures, and the turbulent eddy-viscosity prediction of the Closure Challenge dataset at $Re = 5600$. In the laminar case, FNNs are accurate only within the training regime and with sufficiently informative inputs, whereas GCNs generalize better to unseen geometries, including out-of-distribution flows, and partly compensate for missing input features. In the turbulent proof-of-concept case, the graph-based models achieve modest but consistent improvements compared to FNNs, supporting GCNs as a promising route to non-local RANS closures.
\end{abstract}

\maketitle

\section{Introduction}
The present study is motivated by the classic Reynolds-averaged Navier-Stokes (RANS) equations closure problem \cite{Wilcox}. RANS equations provide a computationally affordable framework for describing the mean behavior of turbulent flows, and are therefore widely employed in both academic and industrial applications involving complex turbulent configurations. However, the averaging operation introduces additional unknowns, represented by the Reynolds stresses, which require suitable closure models \cite{Pope2000}. Classical RANS closures rely on phenomenological assumptions and empirical formulations, whose accuracy and generalizability may be limited when applied to flows that differ from the conditions for which they were developed. For instance, classic linear eddy viscosity models are unable to describe the secondary motion occurring in duct flows, which arises from the anisotropy of the Reynolds normal stresses \cite{Ling2016, oberto2025}.

In the last decade, machine-learning (ML)-based RANS models have attracted increasing attention in the scientific community \cite{Brunton2020, Duraisamy2019, Kutz2017}. Motivated by the limitations of conventional physics-based RANS closures, many data-driven closures have been proposed in recent years \cite{Ling2016, Wang2017, Wu2018, CRUZ2019104258, Rincon_2023, deZordoBanliat2024, Oulghelou2025, Buchanan2026-js}. These models differ from each other in several aspects, including i) the target quantity, namely the anisotropic Reynolds stress tensor \cite{Ling2016} or the Reynolds force vector \cite{CRUZ2019104258, Brener_Cruz_Thompson_Anjos_2021}; ii) the choice of ML model or architecture, such as random forests \cite{VOLPIANI_randomForests}, feed-forward neural networks (FNNs) \cite{Berrone_2022}, Kolmogorov-Arnold networks \cite{Kalia2025-fs}, and Bayesian learning frameworks \cite{Cherroud2022-rn}; iii) the way ML models interact with the RANS system, ranging from complete replacement of conventional RANS closures \cite{Berrone_2022}, to correction of quantities provided by existing models \cite{Wang2017}, or modification of terms inside the PDEs of RANS models \cite{Amarloo2023_bis}; iv) the strategies adopted to enforce physical properties and constraints like frame-coordinate invariance \cite{Ling2016a} or realizability \cite{Ansari2026-iv}. 

However, the majority of existing approaches rely on a local modeling assumption, i.e., the prediction on a given point depends only on the features provided to the ML model at the same point, without incorporating information from neighboring points. Denoting the target quantity $\tau$, this assumption can be expressed as $\tau = f_\theta(x_i)$, being $x_i$ the local feature vector and $f_\theta$ a parametrized function. Such formulation is particularly convenient from an implementation standpoint, as it allows the use of basic ML architecture such as FNNs. However, a local representation may be inadequate for turbulent flows, for which the state of the flow at a given location is influenced by the surrounding flow field.

This motivates the investigation of ML models capable of including non-local effects. A well-established ML architecture that incorporates non-locality is the Convolutional Neural Network (CNN) \cite{Lecun1998-wc, OcarizBorde2021}. In a CNN, a small learnable kernel is slid across the computational domain, and the output at each cell is a weighted sum of the input features in the cell itself and in its neighbors. Each convolutional layer therefore allows neighboring cells to exchange information, and stacking several layers progressively enlarges the region of the flow field that influences the prediction at a given point. However, this operation relies on the regular, Cartesian arrangement of the data, which limits the direct application of CNNs to the unstructured meshes commonly employed in RANS simulations.

To overcome this limitation, Zhou et al. \cite{Zhou2022} introduce the vector-cloud neural network as a non-local modeling framework. Specifically, the vector-cloud neural network constructs a region-to-point mapping from a cloud of feature vectors to the output variable. It is fundamentally a FNN whose inputs are derived through two key operations: (i) pairwise inner products between feature vectors, which render the representation independent of the coordinate frame by encoding only relative orientations and distances, and (ii) an embedding-and-projection step that aggregates scalar features over the cloud, ensuring invariance to the permutation of points. 

In this work, we propose a non-local ML framework based on Graph Neural Networks (GNNs). Rather than explicitly constructing a feature cloud and preprocessing the input data to obtain the desired permutation invariance property, we exploit the graph structure naturally induced by the computational mesh. GNNs have emerged as a powerful ML paradigm for computational sciences, where physical fields are often discretized on unstructured meshes \cite{Pfaff2020-vd, Gladstone2024-yi, Pichi2024-ir, DInverno2025-nx}. In this setting, mesh elements can be represented as graph nodes, while their spatial connectivity defines the graph edges. This representation allows information to be propagated between neighboring points through message-passing operations, thereby providing a natural mechanism to incorporate non-local spatial information into the ML model.

Among GNN architectures, we adopt Graph Convolutional Networks (GCNs) \cite{kipf2017}. GCNs provide a particularly simple and computationally efficient framework for performing such message passing. Similarly to CNNs, by repeatedly aggregating information from neighboring nodes, a GCN allows the prediction at a given mesh point to depend on an increasingly large portion of the surrounding computational domain. However, unlike CNNs, GCNs do not require any Cartesian structure of the underlying computational mesh. Thus, GCNs are a natural choice for investigating the role of non-local information in ML-based RANS modeling.

We present a systematic comparison between FNNs and GCNs for two test flows in the periodic hills geometry \cite{ercoftacPH, Gloerfelt_2019, Xiao2020, Oberto_2025, Ivagnes2025-mu}. The first is a laminar flow at $Re=100$ coupled to a linear advection-diffusion PDE \cite{Zhou2022}, which reproduces some key features of PDEs arising in RANS closures in a simplified setting. We use this case to carry out ablation studies and to investigate specific methodological choices. In particular, we also test a $k$-hop based subgraph sampling for data augmentation of the GCN dataset. The second is the turbulent flow at $Re=5600$ from the Closure Challenge benchmark \cite{closurechallenge}, which we use as a proof-of-concept to show that GCNs are applicable to more challenging cases. For this reason, we deliberately reuse the input features and model setup of the first case, even though they are not tailored to turbulent RANS flows: the goal is not to obtain the most accurate model possible, but to assess the benefit of non-local information within a common framework.

In the laminar case, our results suggest that a FNN model is sufficient to provide accurate results when employed within the training regime and when the input features have sufficient information. As soon as one of these two conditions is not satisfied, FNN performances degrade. In contrast, the GCN model generalizes better on new geometries or unseen flows and better compensates possible lack of information from input features by exploiting non-local information. For the turbulent case, we observe that the GCN model consistently outperforms the FNN one.

Besides this introduction, the paper is organized as follows. Section \ref{sec:rans_section} introduces the RANS closure problem and the local nature of most ML-based RANS models. Section \ref{sec:gnn_section} presents graph convolutional networks and how computational meshes can be viewed as graphs. Section \ref{sec:methodology} describes the two test flows, and Sections \ref{sec:results_transport} and \ref{sec:results_rans} report the corresponding results. Conclusions and future perspectives are given in Section \ref{sec:conclu}.

\vspace{1cm}

\section{RANS equations and their closure problem} \label{sec:rans_section}
In this Section we first present the steady RANS equations and discuss their closure problem. Then, we introduce the main aspects and features of ML-based closure models available in the literature to contextualize the present study.
\subsection{RANS equations}
Steady RANS equations read
\begin{equation} \label{eq:RANS}
\begin{cases}
\nabla \cdot \mathbf{u} = 0, \\
\nabla \cdot (\mathbf{u}\otimes\mathbf{u}) - \nabla \cdot (2\nu \mathbf{S}) = -\nabla p - \nabla \cdot \boldsymbol{\tau},
\end{cases}
\end{equation}
being $\mathbf{u}$ and $p$ the Reynolds-averaged velocity and pressure fields, respectively, $\mathbf{S} = 1/2 \ [\nabla \mathbf{u} + (\nabla \mathbf{u})^T]$ the mean strain rate tensor and $\nu$ the kinematic viscosity of the fluid. Finally, $\boldsymbol{\tau}$ is the Reynolds stress tensor (RST), a symmetric second order tensor arising from the Reynolds-averaging procedure of the nonlinear term of the Navier-Stokes equations \cite{Pope2000}. 

RANS equations \eqref{eq:RANS} are not closed since the unknowns $(\mathbf{u}, p , \boldsymbol{\tau})$ are more than the available equations. To overcome this issue, classic approaches model the RST either algebraically or as solution of PDEs. We refer to \cite{Pope2000, Wilcox} and citations therein for a complete overview on the topic, while we limit our discussion to models based on the Boussinesq's hypothesis \cite{Boussinesq}. Let us define the anisotropic Reynolds stress tensor $\boldsymbol{b} = \boldsymbol{\tau} - 2/3 k \boldsymbol{I}$, being $k = 1/2 \ \text{tr}(\boldsymbol{\tau})$ the turbulent kinetic energy and $\boldsymbol{I}$ the identity tensor. The Boussinesq's hypothesis reads
\begin{equation}\label{eq:Boussinesq}
\boldsymbol{b} = -2 \nu_t \mathbf{S},
\end{equation}
where $\nu_t$ is the turbulent viscosity. Hence, in \eqref{eq:Boussinesq} a linear dependence of $\boldsymbol{b}$ from $\mathbf{S}$ is assumed. Now, the RANS closure problem consists of obtaining an expression for $\nu_t$. Classic and well-established models belonging to this family are the $k-\varepsilon$ \cite{Launder1974}, $k-\omega$ \cite{Wilcox_2008}, $k-\omega$ SST \cite{Menter1994-ii} models. 

Despite the first proposed RANS models have around half a century of history, a universal, reliable model is still missing. This is largely due to the flow-dependent or heuristic assumptions required to derive closure equations, which make each model well-suited to only a limited class of flows and ill-suited to others \cite{NASA}. 

\subsection{Machine-learning-based RANS turbulence models}

The limitations of conventional turbulence closures have motivated the development of data-driven approaches aimed at exploiting the increasing availability of high-fidelity simulations and experimental data. In particular, ML techniques have been investigated to identify functional relationships between local flow features and the unclosed terms appearing in the RANS equations \cite{Duraisamy2019,Duraisamy2021-yn}.

Several strategies have been proposed depending on the quantity to be learned. A first class of approaches retains the structure of an existing turbulence model and uses ML to correct its predictions, for example by learning discrepancies between modeled and high-fidelity Reynolds stresses \cite{Wang2017}. Alternatively, ML models can directly predict the Reynolds stress tensor, its divergence \cite{Berrone_2022, CRUZ2019104258} or, more commonly, its anisotropic component $\boldsymbol{b}$ \cite{Ling2016}. 

Nevertheless, a large fraction of ML-based turbulence closures rely on a pointwise mapping of flow features. Let $x_i \in \mathbb{R}^{d_x}$ denote the vector of input features associated to the $i$-th computational cell, and let $y_i \in \mathbb{R}^{d_y}$ denote the corresponding target quantity. Denoting by $f_\theta$ the parametrized function underlying the ML closure model, the training data consist of pairs ${(x_i,y_i)}_{i=1}^{N}$, and the model is trained to satisfy
\begin{equation}
f_\theta(x_i) \approx y_i,
\qquad i=1,\ldots,N.
\end{equation}
In this formulation, each computational cell is treated independently and the prediction at cell $i$ depends exclusively on the features $x_i$ associated with that cell. Consequently, the spatial connectivity of the computational mesh is not explicitly exploited, and information available in neighboring cells does not directly enter in the prediction.

However, such a pointwise formulation may be restrictive for turbulence modeling \cite{Zhou2022}, as turbulent phenomena inherently involve spatial interactions and the state of the flow at a given location can depend on the surrounding flow field. This motivates the use of ML architectures capable of exploiting the spatial connectivity of the computational mesh and, hence, incorporating non-local information into the closure. This motivates of the present work to investigate the capability of Graph Neural Networks to deal with non-local closures. In the next Section, we define such architecture and describe how naturally a graph structure emerges from computational meshes of CFD simulations.

\vspace{1cm}
\section{Graph Neural Networks}\label{sec:gnn_section}

Many scientific and engineering problems involve data defined on domains that can be naturally represented as graphs. A graph provides a representation in which the entities of the problem are associated with nodes, while their pairwise relationships are encoded through edges. 
In this section, we first introduce graphs and the message-passing framework underlying most modern graph neural networks (GNNs). Then, we describe Graph Convolutional Networks (GCNs) as a particular instance of this framework. We finally discuss how GNNs can be adapted to data arising from CFD simulations. For a comprehensive overview of GNNs and their architectures, we refer the reader to~\cite{Zhou_2020}.

\subsection{Graphs}
In this work, we consider undirected and unweighted graphs, defined as a pair
\begin{equation}
G=(V,E),
\end{equation}
where $V=\{v_1,\ldots,v_{N_v}\}$ is a finite set of nodes and
\begin{equation}
E\subseteq\big\{\{u,v\}:u,v\in V, u \ne v\big\}
\end{equation}
is the set of edges connecting pairs of nodes. Since the graph is undirected, an edge $\{u,v\}$ does not have an associated direction, and therefore $\{u,v\}=\{v,u\}$.

Equivalently, the graph topology can be represented by the adjacency matrix
$A\in\{0,1\}^{N_v\times N_v}$, whose entries are given by
\begin{equation}
A_{ij} =
\begin{cases}
1, & \text{if } \{v_i,v_j\}\in E,\\
0, & \text{otherwise}.
\end{cases}
\end{equation}
Since we consider undirected graphs only, $A$ is symmetric.

Information about the physical or computational quantities associated with the nodes are represented through node features. Specifically, we denote by $x_i \in \mathbb{R}^{d_0}$ the feature vector associated with node $v_i$, where $d_0$ is the number of input features. The collection of all node features can be arranged into the feature matrix
\begin{equation}
X =
\begin{bmatrix}
x_1 \vert \dots \vert x_{N_v}
\end{bmatrix}^T
\in\mathbb{R}^{N_v\times d_0}.
\end{equation}
Depending on the application, these features may represent physical quantities, geometrical information, or other properties associated with the nodes. In the context of CFD, for example, node features may include local flow variables, spatial coordinates, or other quantities defined at the corresponding computational locations.

For an undirected graph, the neighborhood of node $v_i$ is defined as
\begin{equation}
\mathcal{N}(i)
=
\{j:\{v_i,v_j\}\in E\},
\end{equation}
i.e., the set of all nodes that share an edge with $v_i$.

\subsection{Message Passing Framework}\label{sec:gnn}

Most modern GNNs follow the message-passing paradigm, in which the feature representation of each node is iteratively updated by combining its current state with information received from neighboring nodes. Let
$h_i^{(l)}\in\mathbb{R}^{d_l}$
denote the feature representation of node $v_i$ at layer $l$, with
$h_i^{(0)}=x_i$. A general message-passing layer can be expressed as
\begin{equation}\label{eq}
h_i^{(l+1)}
=
\Phi^{(l)}
\left(
h_i^{(l)},
\operatorname{AGG}
\left(
\left\{
\Psi^{(l)}
\left(h_i^{(l)},h_j^{(l)}\right)
:
j\in\mathcal{N}(i)
\right\}
\right)
\right),
\end{equation}
where $\Psi^{(l)}$ is a learnable message function, $\operatorname{AGG}(\cdot)$ is a permutation-invariant aggregation operator, such as summation, averaging, or maximization, and $\Phi^{(l)}$ is a learnable update function. The use of a permutation-invariant aggregation operator ensures that the resulting node representation does not depend on the ordering of the neighboring nodes.

Starting from the initial features $h_i^{(0)}=x_i$, each message-passing layer incorporates information from the neighboring nodes. Consequently, after $k$ layers, the representation of a node can depend on information within its $k$-hop neighborhood. This provides GNNs with a natural mechanism for representing spatially distributed interactions while preserving the underlying graph structure.

\subsection{Graph Convolutional Networks}\label{sec:gnn:gcn}
\begin{figure}[t]
  \centering
  \sbox0{\includegraphics[width=0.8\linewidth]{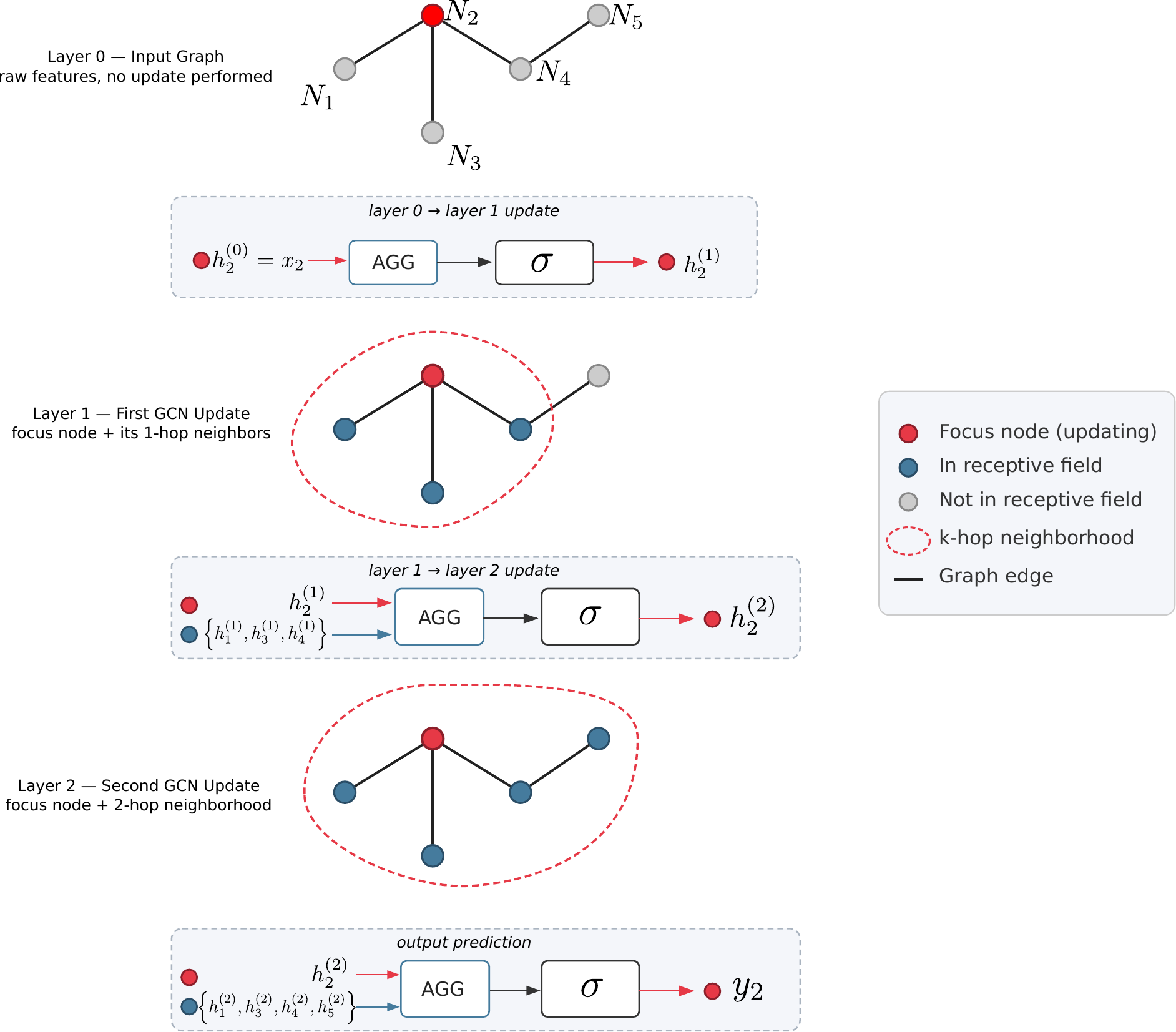}}
  \makebox[0.70\wd0][l]{\usebox0}
  \caption{Graphical representation of GCN propagation rule for node $N_2$. At each layer, the feature vector of every node is updated by i) aggregating the feature vectors of its 1-hop neighbors and its own, with weights given by $W^{(l)}$, and ii) applying a nonlinear function $\sigma$. After $k$ layers, the feature vector of $N_2$ depends on its $k$-hop neighborhood.}
  \label{fig:GCN}
\end{figure}

Among the GNN architectures proposed in the literature, Graph Convolutional Networks (GCNs) provide a computationally efficient way to perform convolution-like operations on graphs. The GCN definition proposed in \cite{kipf2017} rules the propagation of a GCN layer in matrix form as
\begin{equation}\label{eq:gcn}
  H^{(l+1)} =
  \sigma\!\left(
    \tilde{D}^{-1/2}\tilde{A}\tilde{D}^{-1/2} H^{(l)} W^{(l)}
  \right),
\end{equation}
where $H^{(l)} \in \mathbb{R}^{N_v \times d_l}$ is the matrix of node features at layer $l$, $W^{(l)} \in \mathbb{R}^{d_{l}\times d_{l+1}}$ is a trainable weight matrix, and $\sigma(\cdot)$ denotes a non linear activation function. The matrix $\tilde{A} = A + I$ is the adjacency matrix with added self-connections, and $\tilde{D}$ is the associated degree matrix, defined as $\tilde{D}_{ii} = \sum_{j=1}^{N_v} \tilde{A}_{ij}$, $\tilde{D}_{ij} = 0$ if $i \ne j$. 
The normalized adjacency operator
\begin{equation}
\tilde{A}_{\mathrm{norm}}
=
\tilde{D}^{-1/2}
\tilde{A}
\tilde{D}^{-1/2}
\end{equation}
performs a weighted aggregation of the features of each node and its neighbors. Thus, Eq.~\eqref{eq:gcn} can be interpreted as first propagating information over the graph and subsequently applying a learnable linear transformation in the feature space. The symmetric normalization accounts for differences in node degree and helps control the scale of the propagated features. Figure~\ref{fig:GCN} graphically schematizes the above discussion.

From a physical perspective, the operator in Eq.~\eqref{eq:gcn} performs a diffusion-like propagation of information across neighboring nodes of the graph, making GCN architectures particularly suitable for problems involving spatial interactions on computational meshes.

\subsection{Isomorphism between computational meshes and undirected graphs}\label{subsec:graphs}

\begin{figure}[ht!]
    \centering
    \includegraphics[width=0.7\linewidth]{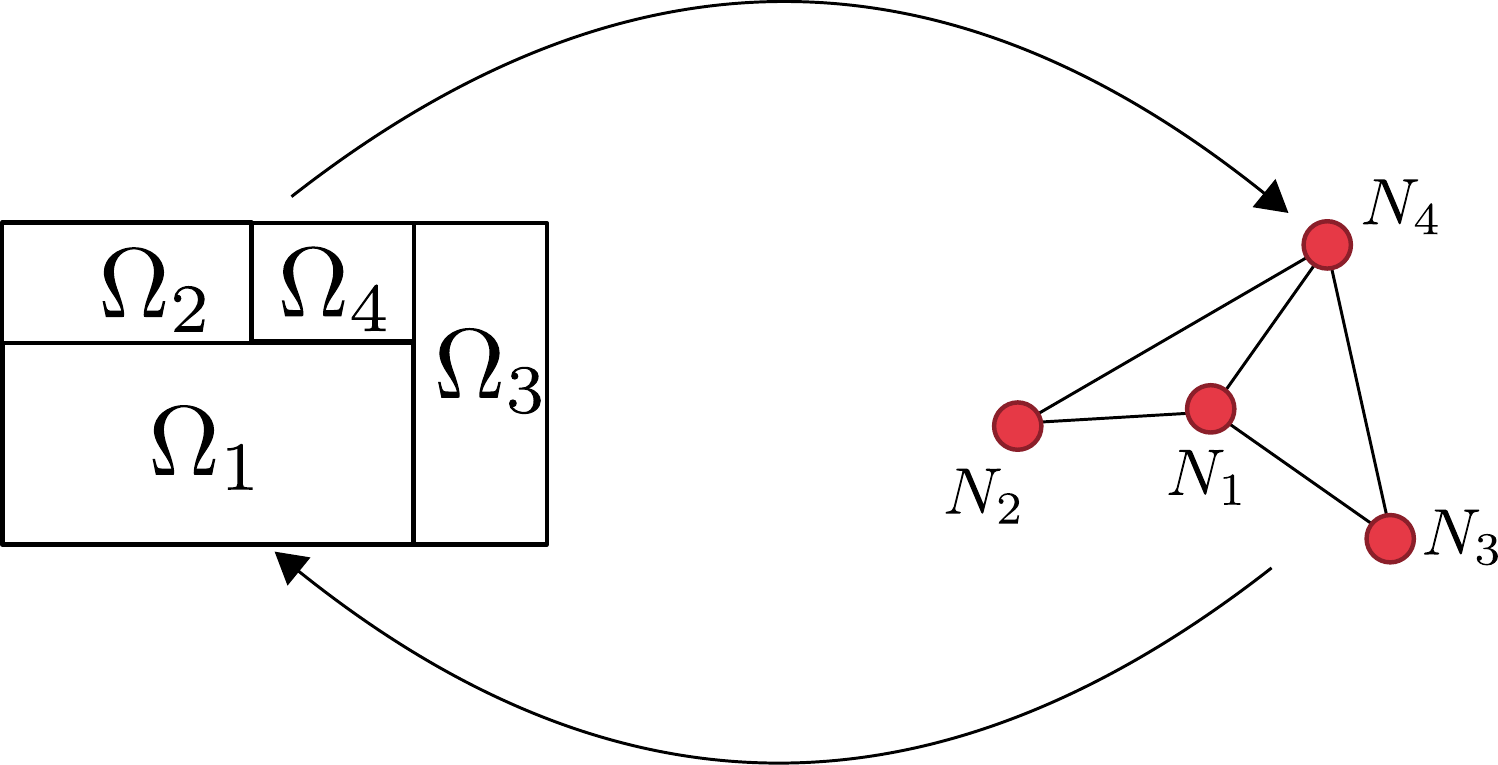}
    \caption{Example of the mesh-to-graph conversion in a finite-volume 2D setting: each mesh cell $\Omega_i$ becomes a graph node $\mathcal{N}_i$, and two nodes are connected by an edge if and only if the corresponding cells share a face.}
    \label{fig:mesh_graph_iso}
\end{figure}

The topology of a computational mesh naturally induces an undirected graph structure, in which mesh entities are represented as graph nodes and their spatial connectivity as graph edges \cite{Du2026-xs}. In the finite-volume setting used in the present work, to each computational cell corresponds a node. Additionally, two nodes are connected if and only if the corresponding cells share a face in the original mesh. This construction is an isomorphism between the mesh's cell-adjacency structure and the resulting graph: the graph carries exactly the connectivity of the mesh, with no topological information lost or added. Fig.~\ref{fig:mesh_graph_iso} illustrates this correspondence on a small mesh patch. We point out that such graph structure naturally arises also for other discretization techniques such as for the finite-element method.

In the presence of periodic boundary conditions, additional edges are added between cells on opposite periodic boundaries, so that information can propagate across the periodic interface exactly as it does physically in the flow. Self-loops are added to every node, so that each node’s own feature vector is included in the aggregation together with those of its neighbors during message passing.

\subsection{Subgraphs sampling strategy}
\label{subsec:sampling}

Classic CFD computational meshes contain thousands or millions of computational cells, making standard GCN training prohibitive \cite{Chiang2019-ud}. To address this issue, in this work we implement a subgraph sampling strategy. Specifically, training samples are generated by extracting smaller graphs from the original mesh graphs. Sampling-based training is well established for large graphs \cite{roznowicz2024largescale,Zhou_2020}, as in GraphSAGE \cite{3294771.3294869}, Cluster-GCN \cite{Chiang2019-ud} and GraphSAINT \cite{Zeng2019-sb}.

In this work, we implement a very simple subsampling strategy: each subgraph is extracted from the full graph, starting from  a  randomly selected seed node, by a breadth-first search (BFS) exploration stopped at a maximum depth $k$. Given the intended simplicity of the algorithm, subgraphs may overlap, and do not cover every node of the original graph.
The resulting subgraphs are $k$-hop neighborhoods of uniform depth, preserving the regular connectivity pattern of the mesh. Under the assumption of a locally regular connectivity with approximately $n$ neighbors per node, a $k$-hop neighborhood contains
\begin{equation}
N \approx 1 + \sum_{i=1}^{k} ni = 1 + \frac{n k(k+1)}{2}
\end{equation}
nodes.  The BFS algorithm is reported in Algorithm~\ref{alg:bfs_sampling}. 

\begin{algorithm}[htbp]
\caption{BFS-based Subgraph Sampling}
\begin{algorithmic}[1]
\Require Graph $G=(V,E)$, maximum depth $K$, number of subgraphs $S$, seed nodes $s$
\Ensure List of sampled graphs $\mathrm{Subgraphs}$
\State $\mathrm{Subgraphs} \gets [\;]$
\State Build adjacency list $\mathrm{neighbors}$ from $E$
\State Randomly shuffle node indices $V$ and select first $S$ as seed nodes
\For{each seed $s$ in seeds}
    \State $\mathrm{visited} \gets \{s\}$
    \State $\mathrm{queue} \gets [(s,0)]$ \Comment{tuple = (node, depth)}
    \While{queue not empty}
        \State Pop $(v,d)$ from queue
        \If{$d \ge K$} 
            \State \textbf{continue}
        \EndIf
        \For{each neighbor $u$ of $v$}
            \If{$u \notin \mathrm{visited}$}
                \State Add $u$ to $\mathrm{visited}$
                \State Append $(u,d+1)$ to queue
            \EndIf
        \EndFor
    \EndWhile
    \State Add graph induced by $\mathrm{visited}$ to $\mathrm{Subgraphs}$
\EndFor
\State \Return $\mathrm{Subgraphs}$
\end{algorithmic}\label{alg:bfs_sampling}
\end{algorithm}

\vspace{1cm}

\section{Investigated flows}\label{sec:methodology}

The methodology is assessed on two benchmarks of increasing complexity: a simplified scalar transport problem on a family of parameterized periodic hills (PH), and the RANS closure problem on the Closure Challenge dataset \cite{closurechallenge}. The two benchmarks share input-feature construction, mesh-to-graph conversion, and subgraph sampling strategy. 

\subsection{Transport problem}\label{sec:transport}

The first benchmark is the scalar transport framework of Zhou et al.~\cite{Zhou2022}. The scalar quantity $\tau$ is defined as the solution of the following advection–diffusion transport equation:

\begin{equation}\label{eq:Zhou_PDE}
\begin{cases}
\mathbf{u} \cdot \nabla \tau - \nabla \cdot ( C_{\nu}\nabla \tau) = \mathrm{P} - \mathrm{E} \\
\mathrm{P} = C_{g} \ell_m s^{2} \sqrt{\tau} \\
\mathrm{E} = C_{\zeta} \sqrt{|\tau|} \tau \\
s = \|\mathbf{S}\| 
\end{cases}
\end{equation}

Here, $\mathbf{u}$ denotes the velocity field and $\mathbf{S}$ the
strain-rate tensor. The mixing length $\ell_m = \min(\kappa\,\hat{\eta},\,C_{\mu}\,\delta)$ is proportional to the wall distance $\hat{\eta}$ and limited by the boundary layer thickness $\delta$. 
The source terms $\mathrm{P}$ and $\mathrm{E}$ represent production and dissipation. The coefficients take values $C_\mu = 0.09$, $\kappa = 0.41$, $C_g = 200\,\mathrm{m}^{-1}\mathrm{s}$, $C_\nu = 0.1\,\mathrm{m}^2/\mathrm{s}$ and $C_\zeta = 3\,\mathrm{s}^{-1}$, related to production, diffusion and dissipation respectively \cite{Zhou2022}. 

Advection and diffusion couple the solution in each cell to its neighborhood: the scalar value at a point depends on the flow state over a surrounding region, not on local quantities alone. Equation~(\ref{eq:Zhou_PDE}) therefore shares the non-local character of turbulent transport, at a much lower numerical cost. This property motivates its use as a first benchmark for graph-based architectures before the RANS closure task. For further details, we refer to \cite{Zhou2022}.

\begin{figure}[ht!]
    \centering
    \includegraphics[width=0.7\linewidth]{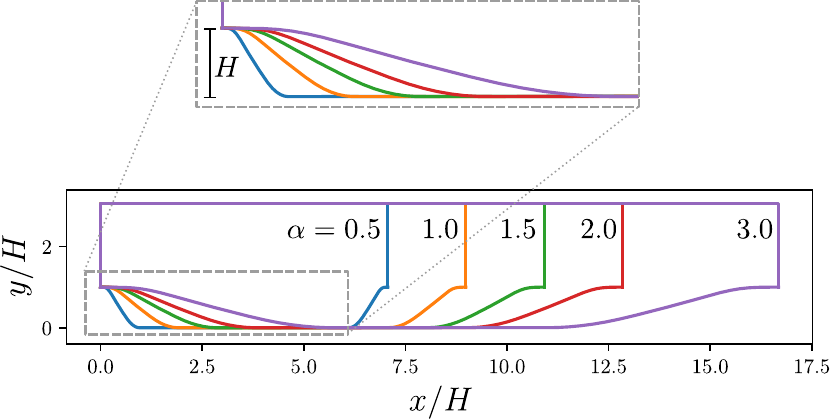}
    \caption{Family of periodic hill geometries obtained by varying the slope parameter $\alpha$. The parameter $\alpha$ controls the steepness of the hill profile while the hill height remains constant. Increasing $\alpha$ produces progressively smoother hill shapes, whereas smaller values of $\alpha$ correspond to steeper geometries.}
    \label{fig:configuration alpha OF_case}
\end{figure}

A family of geometries is generated by the slope parameter $\alpha$, defined as the ratio between the width of the parametrized hill and that of the baseline geometry ($\alpha = 1$). The hill height is held constant; increasing $\alpha$ yields progressively smoother profiles (Fig.~\ref{fig:configuration alpha OF_case}).

Simulations are performed with the open-source software OpenFOAM~\cite{OpenFOAM}. The steady incompressible flow over the periodic-hills family at $Re = 100$ is computed using the SIMPLE pressure--velocity coupling algorithm \cite{Patankar_1972}. The scalar equation~(\ref{eq:Zhou_PDE}) is then solved with a custom solver adapted from the implementation of \cite{Zhou2022}. No-slip conditions ($\mathbf{u} = \mathbf{0}$) are imposed on the walls, with Dirichlet condition $\tau = 0$; cyclic conditions at inlet and outlet enforce domain periodicity.

Equation~\eqref{eq:Zhou_PDE} is dimensional; the flow features extracted to define the inputs of the investigated models are set dimensionless, as reported in Tab.~\ref{tab:input_features}. Specifically, they are obtained from the computational mesh and normalized by fixed reference scales: the velocity magnitude by $U_{\text{ref}}$, the strain-rate and vorticity magnitudes, the latter denoted by $\vert \vert \boldsymbol{W} \vert \vert $, by the reference time $t_{\text{ref}} = 1$, and the geometric quantities (wall distance and mixing length) by the reference lengths $L_{\text{ref}}$ and $\Delta_{\text{ref}} = C_\mu \delta$. The reference values are $L_{\text{ref}} = 1.518$, $U_{\text{ref}} = 1.387$ and $\Delta_{\text{ref}} = 0.135$.

\begin{table}[h!]
\centering
\caption{Input features used for ML-based models.}
\label{tab:input_features}
\begin{tabular}{ll}
\hline
\textbf{Input} & \textbf{Definition} \\
\hline
\\ 
Velocity magnitude & $|\mathbf{u}| / U_{\text{ref}}$ \vspace{0.1cm} \\
Strain-rate magnitude & $\|\mathbf{S}\| / t_{\text{ref}}$ \vspace{0.1cm} \\
Vorticity magnitude & $\|\boldsymbol{W}\| / t_{\text{ref}}$ \vspace{0.1cm} \\
Wall distance & $\hat{\eta} / L_{\text{ref}}$ \vspace{0.1cm} \\
Mixing length & $\ell_m / \Delta_{\text{ref}}$ \vspace{0.1cm} \\
\hline
\end{tabular}
\end{table}

\subsection{RANS closure on the Closure Challenge periodic hills}\label{sec:rans}
The second benchmark addresses the RANS closure problem directly. 
The target quantity is the turbulent eddy viscosity $\nu_t$, learned from the PH database of the Closure Challenge benchmark~\cite{closurechallenge} at $Re = 5600$. The database provides 29 DNS simulations obtained by varying the hill shape and the domain geometry. Each case is identified by the label $\texttt{alpha\_XX\_YYYY\_ZZZZ}$: $\mathrm{XX}$ denotes the hill shape parameter $\alpha$, $\mathrm{YYYY}$ the streamwise domain length $L_x/H$, $\mathrm{ZZZZ}$ the domain height $L_y/H$. Training, validation and testing follow the predefined split of the benchmark protocol \cite{closurechallenge}, with the test set comprising unseen configurations featuring geometries that are substantially different from those represented in the training set.

Four of the five flow features of the transport framework (Tab.~\ref{tab:input_features}) are computed from the mean fields of each RANS configuration: the velocity, strain-rate and vorticity magnitudes, and the wall distance. The mixing length is excluded. In the RANS solver, $\ell_m$ is a closed-form function of the eddy viscosity and the strain-rate magnitude, $\ell_m = (\nu_t / (\sqrt{2}\,\|\boldsymbol{S}\|))^{1/2}$
. Since $\|\boldsymbol{S}\|$ is itself an input feature, including $\ell_m$ would give the models direct algebraic access to the regression target through two of their own inputs. The effect of this choice on the results is quantified in Appendix~\ref{app:gcn_rans_training}.

An alternative target, obtained by projecting the DNS Reynolds stresses onto the mean strain-rate tensor \cite{Wu2019, Brener_Cruz_Thompson_Anjos_2021},
\begin{equation*}
\nu_t = -\frac{1}{2}\, \frac{\boldsymbol{\tau} : \mathbf{S}}
                          {\mathbf{S} : \mathbf{S}},
\end{equation*}
was also tested.
The projected field is affected by significant numerical noise: the operation involves the mean velocity gradients, which amplify the noise of the averaged fields, and a division by $\mathbf{S}:\mathbf{S}$, which approaches zero in regions of weak mean strain, where small errors in the DNS Reynolds stresses are strongly amplified and spurious large or negative values appear. The corresponding learning task yields poor accuracy \cite{Wu2019}. The RANS-provided eddy viscosity is therefore retained as target, being a more stable and consistent quantity.

It is worth noting that neither the input variables \cite{Ling2016, oberto2025} nor the target quantity adopted in this second test represent the usual choices for a RANS-based data-driven closure. The same input variables as in the previously investigated flow are retained for consistency and ease of comparison, avoiding the introduction of a new input framework at this stage. This choice is motivated by the proof-of-concept nature of this experiment, whose primary purpose is to further assess the proposed graph-based methodology in a RANS setting rather than to define a final, high-fidelity turbulence-closure model.

\vspace{1cm}
\section{Results: transport phenomena}\label{sec:results_transport}

\subsection{Models hyperparameters selection and training strategies}\label{sec: hyperparameters and training}

Three models are compared on the scalar transport benchmark: \texttt{FNN}, \texttt{GCN}, and \texttt{GCN\_khop}. All models have the same number of trainable parameters. \texttt{GCN} and \texttt{GCN\_khop} share the same graph-based architecture; they differ only in the graphs used during training. Specifically, the \texttt{GCN} model is trained on graphs corresponding to the full meshes. Conversely, the \texttt{GCN\_khop} model is trained on sampled subgraphs from the full graphs generated through the $k$-hop BFS sampling strategy. 

We generate simulations for $\alpha \in [1,2]$ with step $0.05$, for a total of 21 simulations. Among them, 4 are randomly selected for validation purposes and the remaining ones are used for training. The test set covers $\alpha \in [0.5,3]$ with step $0.025$, so that the models are evaluated both within and outside the training interval.

\begin{figure}[ht!]
    \centering
    \begin{subfigure}{0.47\textwidth}
        \centering
        \includegraphics[width=\textwidth]{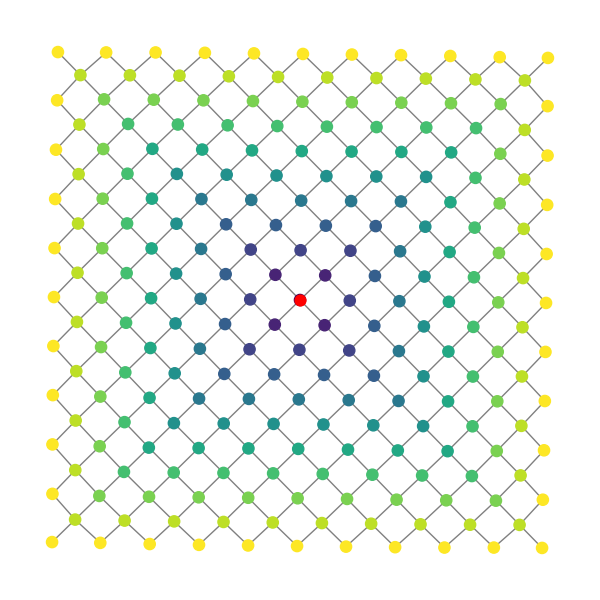}
        \caption{}
    \end{subfigure}
    \hfill
    \begin{subfigure}{0.47\textwidth}
        \includegraphics[width=\linewidth]{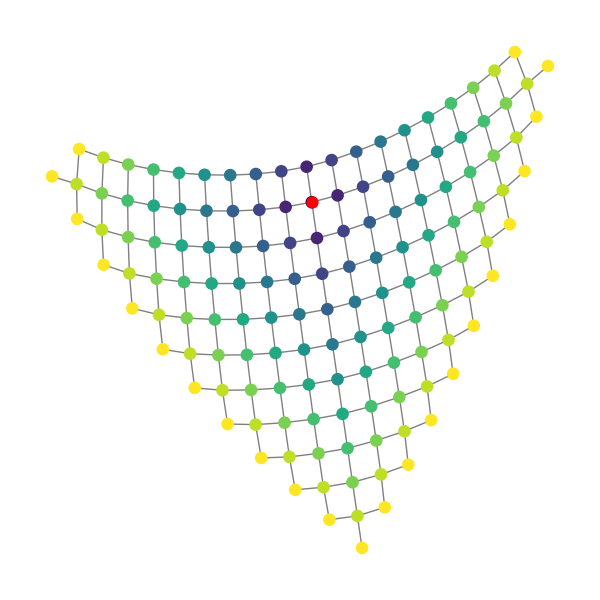}
        \caption{}
    \end{subfigure}
    \caption{Examples of sampled subgraphs extracted from the computational mesh using the $k$-hop sampling algorithm. The starting central red node is extracted randomly, $k$ determines the depth of the graph while the strategy preserves the regular connectivity pattern of the mesh. Nodes are colored depending on their distance from the starting node. The graph on the right corresponds to a selected seed cell close to the upper wall.}\label{fig:sampling_graphs}
\end{figure}

For each training simulation, 400 subgraphs are sampled via the $k$-hop procedure of Section~\ref{subsec:sampling}, for a total of 6800 extracted subgraphs, to train the \texttt{GCN\_khop} model. The value of 400 subgraphs per configuration is the result of a preliminary empirical study, where a good compromise between the increased dataset size and the increased training cost has been found. Figure \ref{fig:sampling_graphs} shows two examples of subsampled graphs used during the training, one of them being close to a wall. We point out that for this flow and the next one, the geometries are simple enough to generate fully quadrilateral meshes. Hence, a CNN model could have been used instead of the more general GCN framework. However, the presented GCN approach is immediately usable for generic meshes, while the CNN one is not.

Once the \texttt{GCN\_khop} is trained using the sampled subgraphs, it performs inference on the full graph, therefore reconstructing the field over the entire domain in a single forward pass. In the following sections, if not explicitly stated, we always select subgraphs with the $k$-hop algorithm with $k=10$.

On the transport benchmark, four parameter budgets are evaluated. They are obtained by varying the width (16, 34, 50, 56), at a fixed depth of 5 hidden layers. 
The budget of 10551 parameters, with 50 neurons per hidden layer for the \texttt{FNN} and 50 channels for the \texttt{GCN}, offers the best trade-off between accuracy and training cost and is retained for both benchmarks. Because $\tau \in [0,1]$ in the investigated flow, we apply a sigmoid function to the output layer of all models.

All models are trained by minimizing the mean squared error with the Adam optimizer, at an initial learning rate of $10^{-3}$. A \texttt{ReduceLROnPlateau} schedule (factor 0.3, minimum learning rate $10^{-6}$, patience 40 epochs) is adopted; a \texttt{OneCycle} schedule was tested and discarded, as it produced less stable training behavior. \texttt{ReduceLROnPlateau} counts its patience in epochs and only reduces the learning rate; training itself is halted by a separate early-stopping criterion, triggered after twice the scheduler patience of Table~\ref{tab:training_config} without improvement in the validation loss. Since the number of updates performed in each epoch differs substantially between models, a full-graph batch and a sampled-subgraph batch process very different numbers of graphs per update, the same nominal patience can correspond to very different numbers of optimization steps across models. This asymmetry is most significant for \texttt{GCN} on the RANS dataset, where it is investigated further in Sec.~\ref{sec:rans_hyperparams} and Appendices~\ref{app:gcn_rans_training}.

The \texttt{FNN} is trained on batches of 1024 cells. The \texttt{GCN} is trained on batches of five full meshes 
while \texttt{GCN\_khop} is trained on batches of 32 sampled subgraphs. The maximum number of training epochs is set to 6000. 

\begin{table}[h!]
\centering
\caption{Final training configuration for the three models. RANS-specific values in parentheses.}
\label{tab:training_config}
\begin{tabular}{llll}
\hline
\textbf{Parameter} & \textbf{FNN} & \textbf{GCN} & \textbf{GCN $k$-hop} \\
\hline
Optimizer & Adam & Adam & Adam \\
Initial learning rate & $10^{-3}$ & $10^{-3}$ & $10^{-3}$ \\
Scheduler & ReduceLROnPlateau & ReduceLROnPlateau & ReduceLROnPlateau \\
Scheduler factor & 0.3 & 0.3 & 0.3 \\
Scheduler patience [epochs] & 40 & 40 (400) & 40 \\
Minimum learning rate & $10^{-6}$ & $10^{-6}$ & $10^{-6}$ \\
Batch size & 1024 & 5 (2) & 32 \\
Maximum epochs & 6000 & 6000 & 6000 \\
\hline
\end{tabular}
\end{table}

\subsection{Comparative analysis: 5 inputs NNs}
\label{sec:5inputs}
The three models are trained on the PHs transport dataset introduced in Section~\ref{sec:transport}: the training set covers $\alpha \in [1,2]$ with step $0.05$, and the trained models are evaluated on the wider range $\alpha \in [0.5,3]$ with step $0.025$. All three models use the five input features of Tab.~\ref{tab:input_features}.
 
\begin{figure}[ht!]
    \centering
    \includegraphics[width=0.9\linewidth]{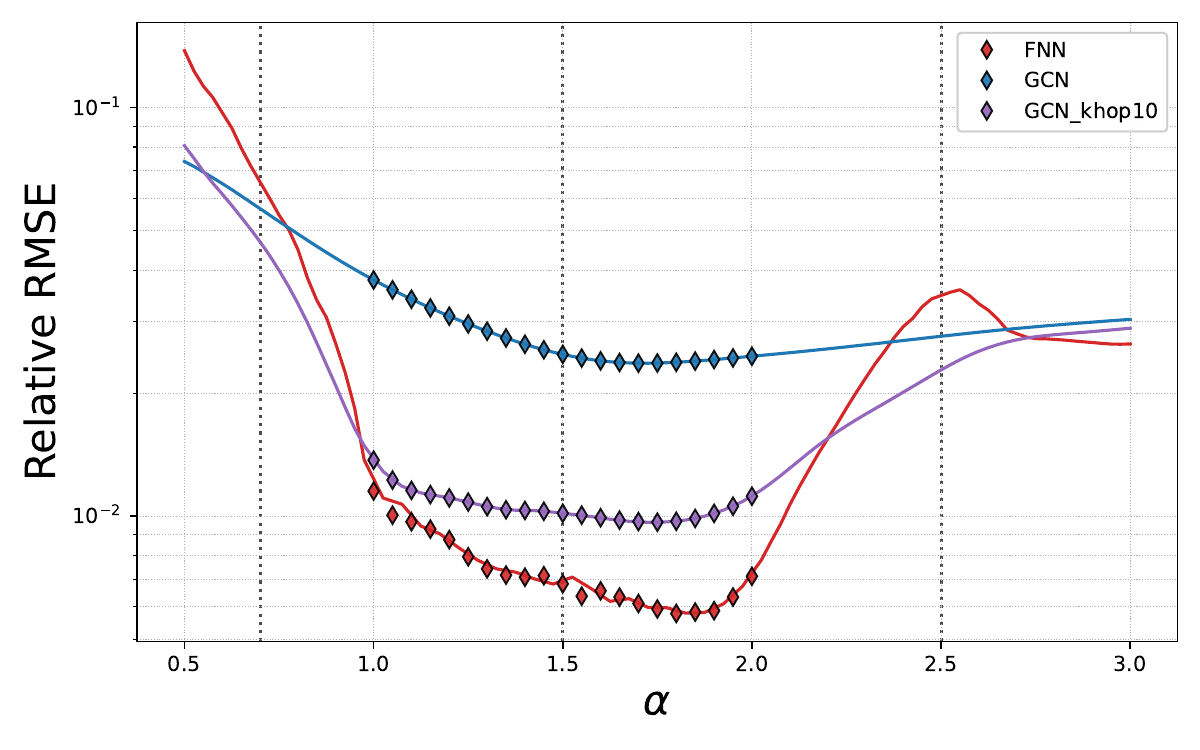}
    \caption{RRMSE as a function of $\alpha$ for FNN, GCN and GCN\_khop. Dashed vertical lines mark $\alpha = 0.7, 1.5, 2.5$, corresponding to the qualitative comparisons of Fig.~\ref{fig:afterthesis_abs_errpredictionT_0.7}, \ref{fig:afterthesis_abs_errpredictionT_1.5} and
\ref{fig:afterthesis_abs_errpredictionT_2.5}.}
    \label{fig:GCN_FNN_10551tp_input5_check}
\end{figure}

Fig.~\ref{fig:GCN_FNN_10551tp_input5_check} reports the relative root mean square error (RRMSE) in the $\tau$ prediction of the three models with respect to the reference solution. The relative RMSE is defined as
\begin{equation}
  \mathrm{RRMSE} = \frac{\displaystyle\sqrt{\sum_{i=1}^{N}
  \left(\tau_i^{\mathrm{pred}} - \tau_i^{\mathrm{sim}}\right)^2}}
  {\displaystyle\sqrt{\sum_{i=1}^{N} \left(\tau_i^{\mathrm{sim}}\right)^2}} ,
  \label{eq:relrmse}
\end{equation}
where $N$ is the number of cells, $\tau_i^{\mathrm{pred}}$ is the value predicted by the model at the $i$-th cell, and $\tau_i^{\mathrm{sim}}$ is the corresponding reference value from the OpenFOAM solution of the transport equation. Diamond symbols are used for training/validation $\alpha$ values. Within the training range, \texttt{FNN} attains the lowest error, while \texttt{GCN\_khop} achieves a lower error than \texttt{GCN}, close to the \texttt{FNN} one. Outside the training interval, the trend reverses. In the lower extrapolation region, $0.5 \le \alpha < 1$, \texttt{GCN\_khop} is the most accurate model, with an average error equal to 0.04321, followed by \texttt{GCN} at 0.05470 and \texttt{FNN} at 0.06381. In the upper extrapolation region, $2 < \alpha \le 3$, \texttt{GCN\_khop} again performs best, with an average error of 0.02182, closely followed by \texttt{FNN} at 0.02456, while \texttt{GCN} is the least accurate, at 0.02752. The error of \texttt{FNN} increases markedly for $\alpha > 2$. The error of \texttt{GCN\_khop}, although consistently lower than the \texttt{GCN}'s one within the training range, tends to converge toward the behavior of \texttt{GCN} as $\alpha$ departs further from the training interval. 
These trends are summarized in Table~\ref{tab:GNN_FNN_10551tp_5input_check}, which also reports the validation loss and the number of training epochs. In the Table, we denote by \textbf{err\_testing\_1} the average error in the interval $\alpha \in [0.5, 1)$ and by \textbf{err\_testing\_2} the average error in the interval $\alpha \in (2, 3]$. \texttt{FNN} attains the lowest validation loss, equal to $7.394\times10^{-6}$, and converges in fewer epochs, equal to 1709, than the two graph-based models, both of which reach the maximum epoch budget of 6000.

\begin{table}[ht!]
    \centering
    \caption{Model comparison for the Transport flow. }
    
    \begin{tabular}{lccccc}
    \hline
    \textbf{Model} & \textbf{Epochs} & \textbf{Val\_loss} & \textbf{err\_testing\_1} &\textbf{err\_training} & \textbf{err\_testing\_2}  \\
    \hline
    FNN & 1709 & 7.394e-06 & 0.06381 & 0.00738 & 0.02456 \\
    GCN & 6000 & 8.062e-05 & 0.05470 & 0.02708 & 0.02752 \\
    GCN\_khop & 6000 & 6.163e-05 & 0.04321 & 0.01051 & 0.02182 \\
    \hline 
    \end{tabular}
    \label{tab:GNN_FNN_10551tp_5input_check}
\end{table}

\begin{figure}[h!]
    \centering
    \begin{subfigure}{0.48\textwidth}
        \centering
        \includegraphics[width=\textwidth]{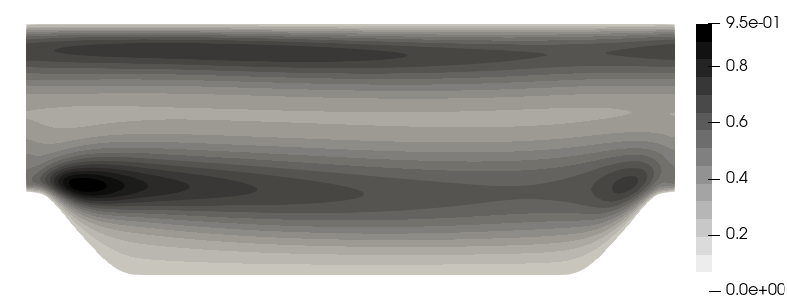}
        \caption{$\tau$}
    \end{subfigure}
    \hfill
    \begin{subfigure}{0.48\textwidth}
        \includegraphics[width=\linewidth]{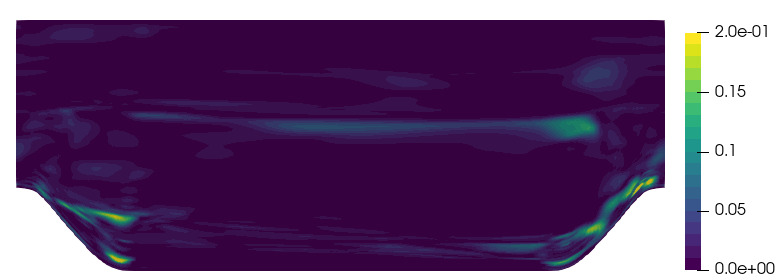}
        \caption{FNN}
    \end{subfigure} 
   
    \begin{subfigure}{0.48\textwidth}
        \includegraphics[width=\linewidth]{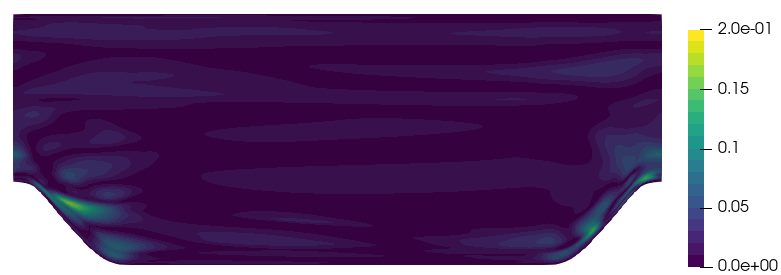}
        \caption{GCN}
    \end{subfigure}
    \hfill
    \begin{subfigure}{0.48\textwidth}
        \includegraphics[width=\linewidth]{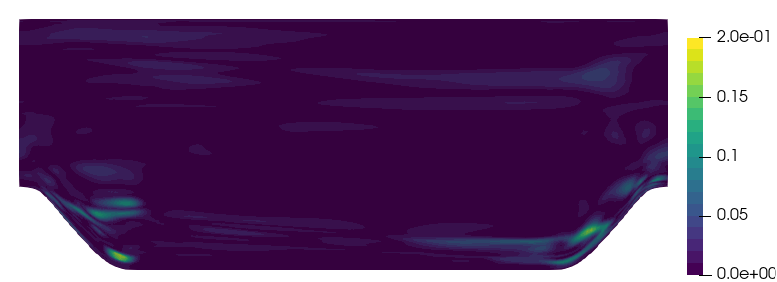}
        \caption{GCN\_khop}
    \end{subfigure}

    \caption{Absolute error of the scalar field $\tau$ for $\alpha = 0.7$:
  (\textit{a}) reference field $\tau$; (\textit{b}) FNN; (\textit{c}) GCN; (\textit{d}) GCN\_khop ($k=10$).} \label{fig:afterthesis_abs_errpredictionT_0.7}
\end{figure}

\begin{figure}[h!]
    \centering
    \begin{subfigure}{0.48\textwidth}
        \centering
        \includegraphics[width=\textwidth]{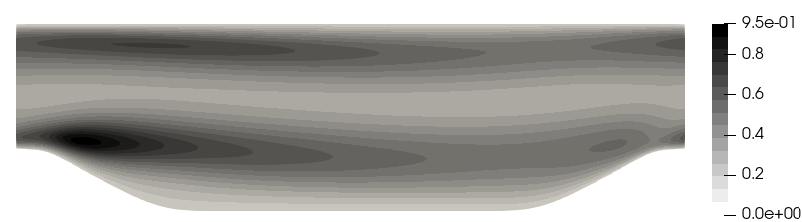}
        \caption{$\tau$}
    \end{subfigure}
    \hfill
    \begin{subfigure}{0.48\textwidth}
        \includegraphics[width=\linewidth]{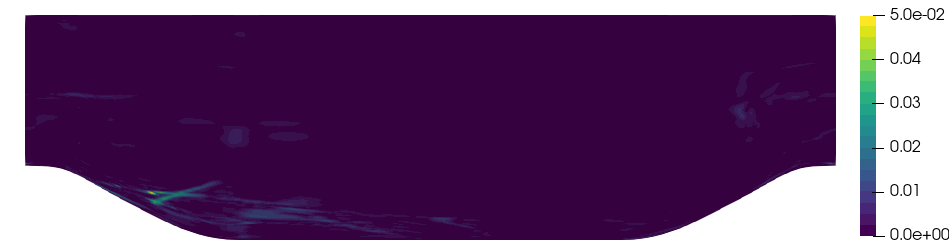}
        \caption{FNN}
    \end{subfigure} 
   
    \begin{subfigure}{0.48\textwidth}
        \includegraphics[width=\linewidth]{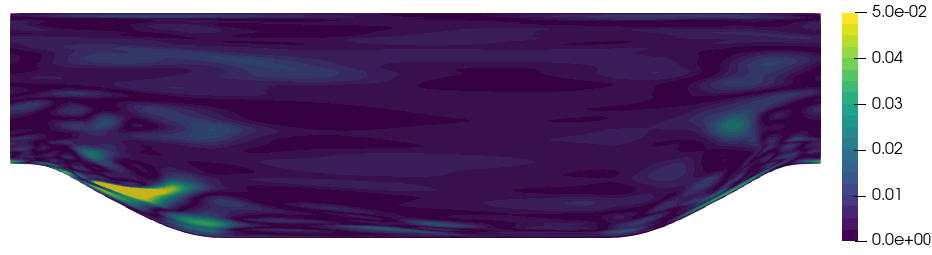}
        \caption{GCN}
    \end{subfigure}
    \hfill
    \begin{subfigure}{0.48\textwidth}
        \includegraphics[width=\linewidth]{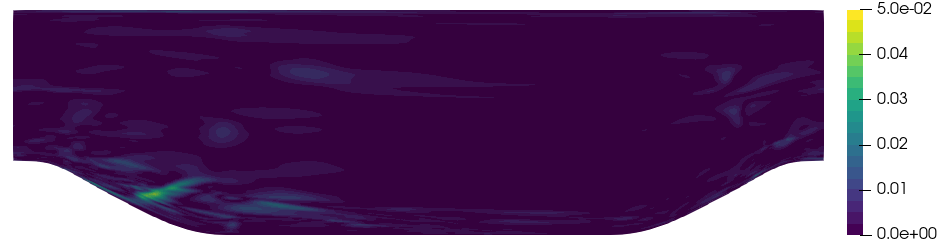}
        \caption{GCN\_khop}
    \end{subfigure}

    \caption{Absolute error of the scalar field $\tau$ for $\alpha = 1.5$:
  (\textit{a}) reference field $\tau$; (\textit{b}) FNN; (\textit{c}) GCN; (\textit{d}) GCN\_khop ($k=10$).} \label{fig:afterthesis_abs_errpredictionT_1.5}
\end{figure}

\begin{figure}[h!]
    \centering
    \begin{subfigure}{0.48\textwidth}
        \centering
        \includegraphics[width=\textwidth]{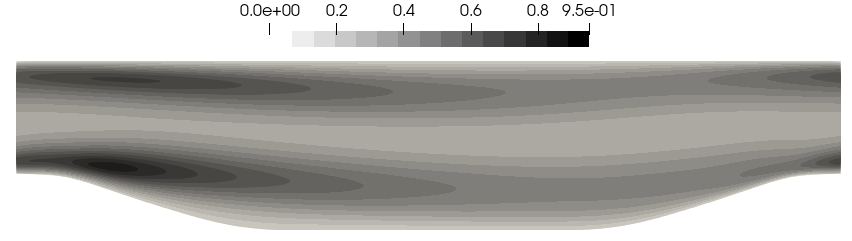}
        \caption{$\tau$}
    \end{subfigure}
    \hfill
    \begin{subfigure}{0.48\textwidth}
        \includegraphics[width=\linewidth]{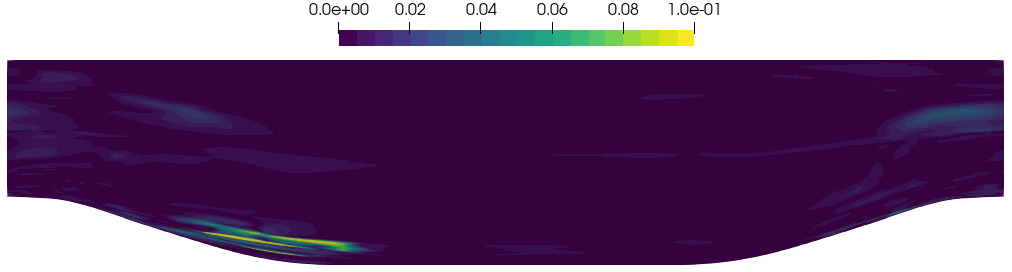}
        \caption{FNN}
    \end{subfigure} 
   
    \begin{subfigure}{0.48\textwidth}
        \includegraphics[width=\linewidth]{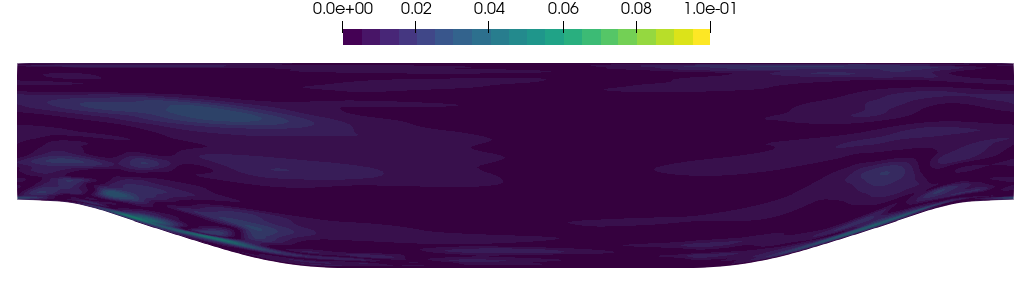}
        \caption{GCN}
    \end{subfigure}
    \hfill
    \begin{subfigure}{0.48\textwidth}
        \includegraphics[width=\linewidth]{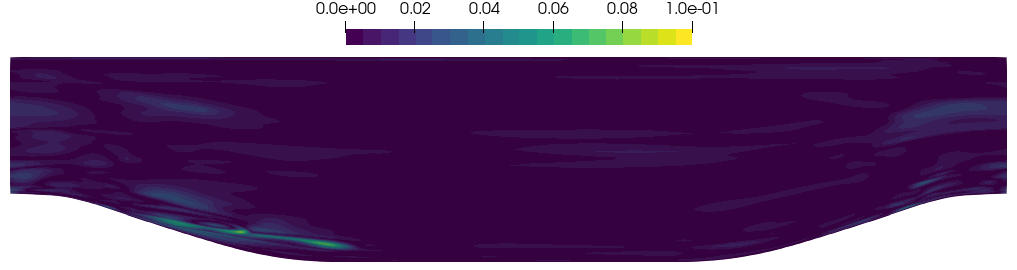}
        \caption{GCN\_khop}
    \end{subfigure}

    \caption{Absolute error of the scalar field $\tau$ for $\alpha = 2.5$:
  (\textit{a}) reference field $\tau$; (\textit{b}) FNN; (\textit{c}) GCN; (\textit{d}) GCN\_khop ($k=10$).} \label{fig:afterthesis_abs_errpredictionT_2.5}
\end{figure}

Figures~\ref{fig:afterthesis_abs_errpredictionT_0.7}, \ref{fig:afterthesis_abs_errpredictionT_1.5} and
\ref{fig:afterthesis_abs_errpredictionT_2.5} show the absolute error of the predicted $\tau$ field for $\alpha = 0.7$, $\alpha = 1.5$ and $\alpha = 2.5$, corresponding to the lower extrapolation, the training, and the upper extrapolation intervals, respectively. \texttt{FNN} shows the largest localized errors, concentrated near the recirculation region at $\alpha = 0.7$ and near the hill crest at $\alpha = 2.5$, with the error increasing markedly between the two cases, consistent with the increase of its error for $\alpha > 2$ discussed above. Within the training range, at $\alpha = 1.5$, the error of \texttt{FNN} remains low and confined to a narrow region near the hill crest. \texttt{GCN} and \texttt{GCN\_khop} show smaller and more uniformly distributed errors than \texttt{FNN} at $\alpha = 0.7$ and $\alpha = 2.5$, and comparably low errors, distributed more uniformly across the domain, at $\alpha = 1.5$, consistent with the quantitative results discussed above.

\newpage
\subsection{Sampling Strategy: Speedup and Accuracy test}\label{sec:sampling_speedup}
\begin{figure}[ht!]
    \centering
    \includegraphics[width=0.9\linewidth]{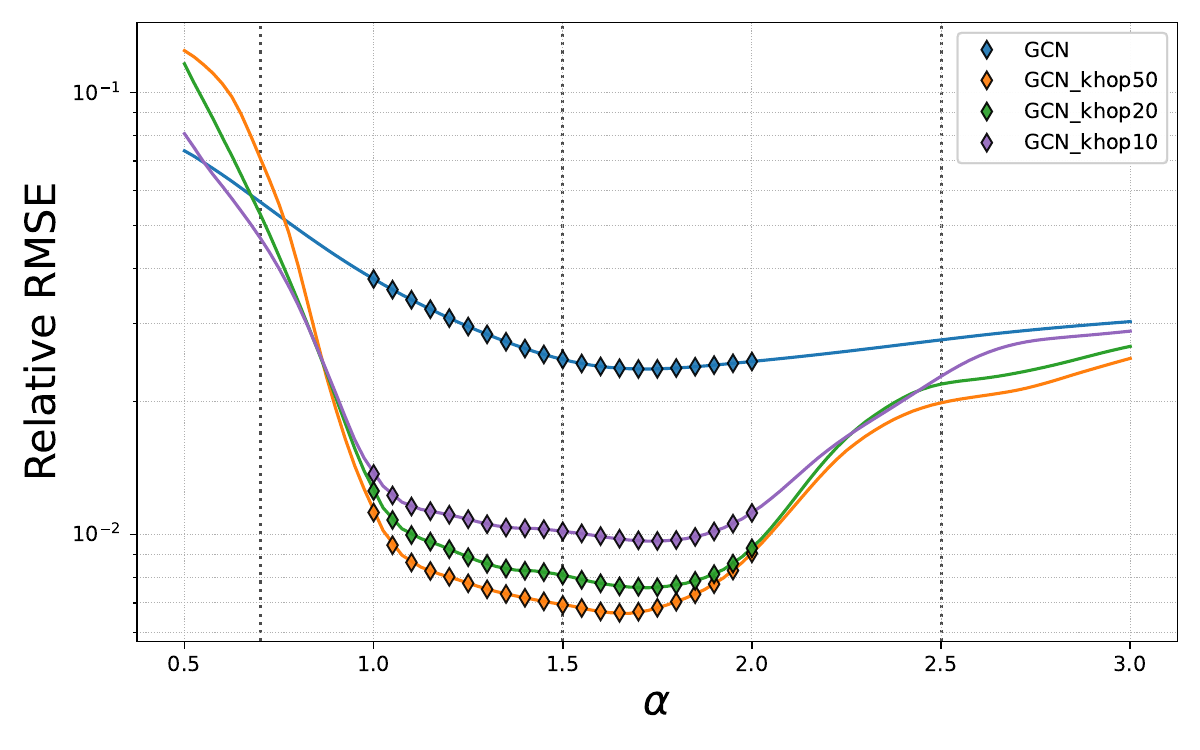}
    \caption{RRMSE as a function of $\alpha$ for \texttt{GCN} and for \texttt{GCN\_khop} with $k$ = 10, 20, 50.}
    \label{fig:GCN_khop_10551tp_check}
\end{figure}

As stated in Sec.~\ref{sec: hyperparameters and training}, \texttt{GCN\_khop} is trained on 400 subgraphs sampled per training configuration ($k$-hop procedure of Sec.~\ref{subsec:sampling}). 
Figure~\ref{fig:GCN_khop_10551tp_check} compares \texttt{GCN} with \texttt{GCN\_khop} for $k=10, 20, 50$, over the full range of $\alpha$. Within and close to the training interval, all three sampled configurations outperform \texttt{GCN}, with $k=20$ and $k=50$  yielding the lowest errors. Outside the training range, in particular for $\alpha > 2.5$, the error of the sampled configurations increases and approaches the \texttt{GCN} one. Conversely, for $\alpha < 1.0$, all \texttt{GCN\_khop} errors quickly increase, reaching higher values than the \texttt{GCN}'s one for $\alpha=0.5$. Specifically, for $\alpha=0.5$ the error ordering is the opposite of that in the training regime. 

This behavior suggests that the benefit in accuracy of $k$-hop sampling is closely related to the similarity between the local graph structures represented during training and those encountered at inference. Within and close to the training interval, subgraph sampling exposes the model to a larger variety of local graph configurations across the training geometries, which may improve its ability to generalize across the range of $\alpha$. As the geometry moves farther from the training range, however, this advantage progressively decreases. This effect is particularly pronounced for $\alpha<1$. This asymmetric behavior in the $\alpha$ intervals indicates that extrapolation beyond the training interval is not equivalent in the two directions and that the effectiveness of local subgraph training depends on the extent to which the resulting local graph and flow structures remain represented by the training data.

\begin{figure}
    \centering
    \includegraphics[width=0.9\linewidth]{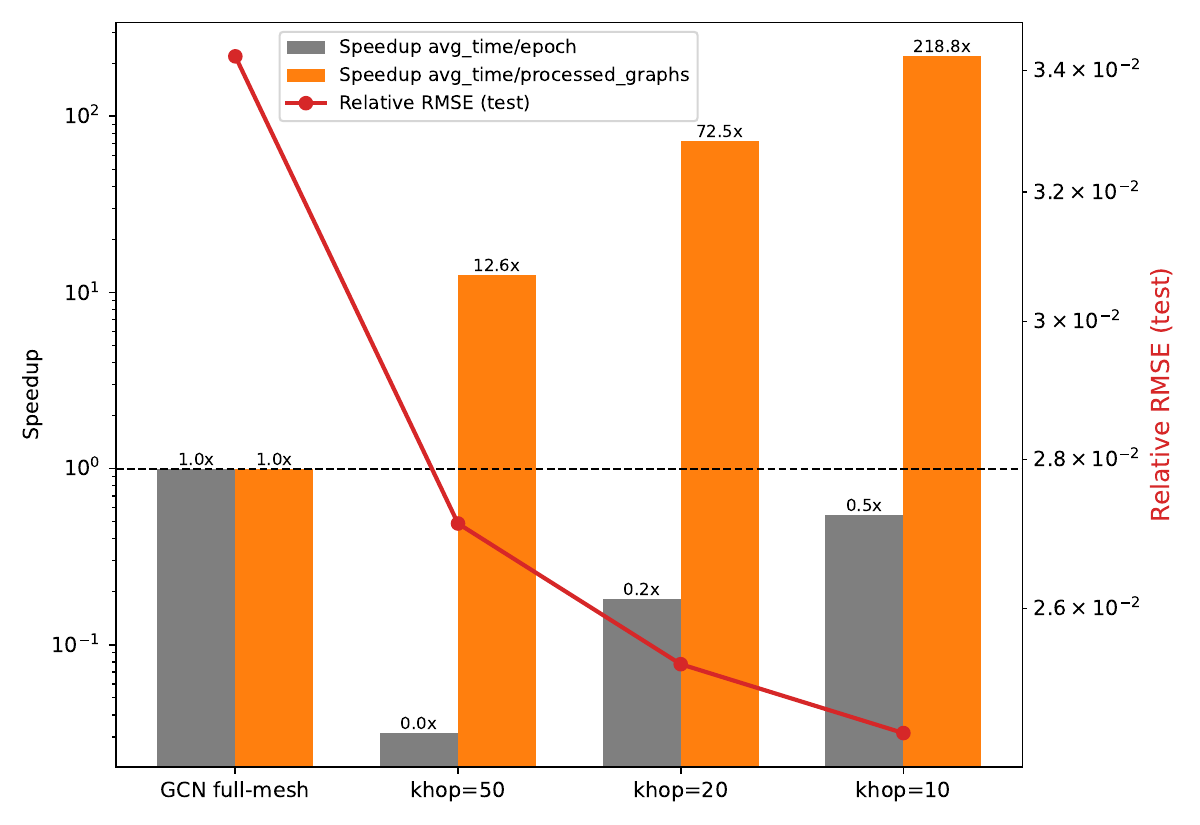}
    \caption{Speedup of the $k$-hop sampling strategy relative to \texttt{GCN}, in terms of average time per epoch and average time per processed graph, together with the RMSE on the test set.}
    \label{fig:GCNkhop_speedup_accuracy_test}
\end{figure}

Fig.~\ref{fig:GCNkhop_speedup_accuracy_test} reports both the training speedup per epoch and the speedup per processed graph with respect to \texttt{GCN}, together with the RRMSE on the test set. As expected, the speedup per processed graph increases markedly as $k$ decreases: from a factor of 12.6 at $k=50$, to a factor of 218.8 at $k=10$. The speedup per training epoch is instead below unity for all sampled configurations, equal to 0.03, 0.18 and 0.54 for $k=50, 20, 10$ respectively, since each epoch processes a significantly larger number of smaller subgraphs instead of a unique large graph. The RRMSE on the test set decreases monotonically as $k$ decreases, from a value of approximately $3.4\times10^{-2}$ for \texttt{GCN} to approximately $2.6\times10^{-2}$ for $k=10$. Therefore, the $k=10$ configuration offers the best combination of accuracy on unseen configurations and per-graph computational cost among the tested configurations. Hence, we adopt $k=10$ as the reference configuration.

\newpage
\subsection{Comparative analysis: Decreasing the number of inputs}\label{sec:decreasing_inputs}
In this Section, we compare models' accuracy when trained with fewer input features. Specifically, we reduce the input feature vector size from five to four, three, two and one. This allows us to investigate the models capabilities when only sparse information is available. Starting from the five features listed in Table~\ref{tab:input_features}, inputs are removed sequentially, in order: mixing length, wall distance, vorticity magnitude, and strain-rate magnitude; leaving velocity magnitude alone in the one-input configuration. 
 
\begin{figure}[ht!]
    \centering
    \begin{subfigure}{0.48\textwidth}
        \centering
        \includegraphics[width=\textwidth]{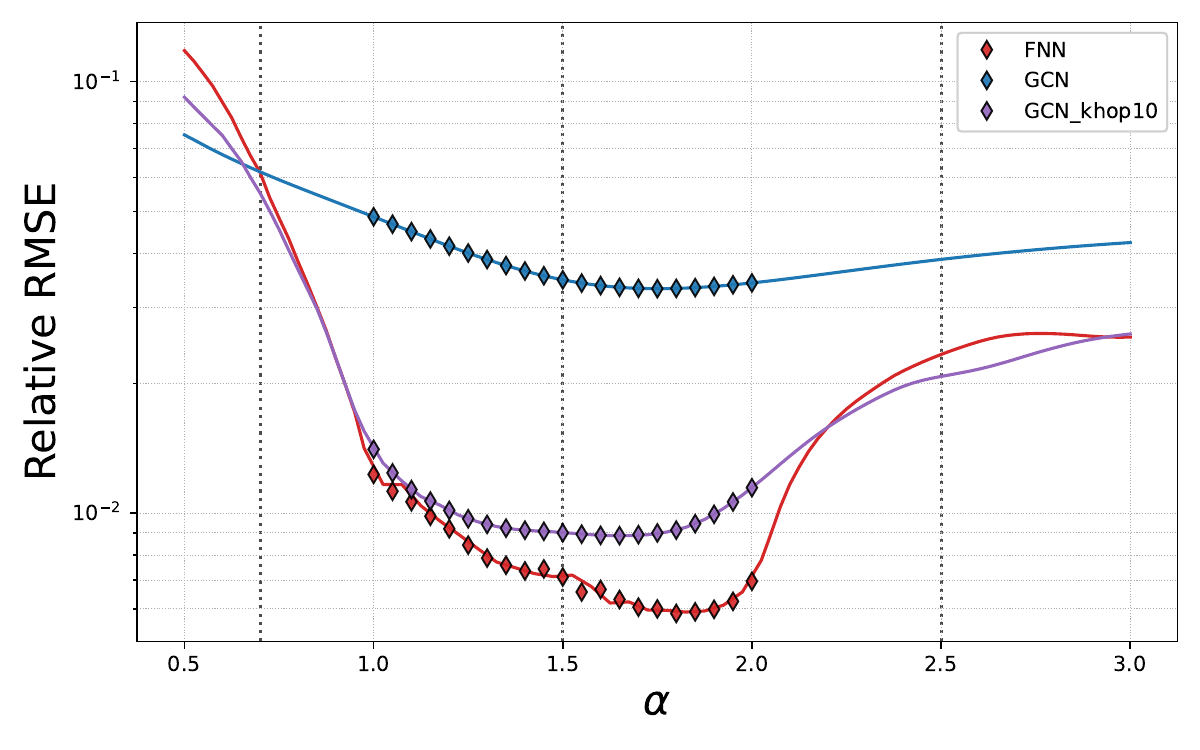}
        \caption{4 inputs}
        \label{fig:error_FNN&GNN_4input}
    \end{subfigure}
    \hfill
    \begin{subfigure}{0.47\textwidth}
        \includegraphics[width=\linewidth]{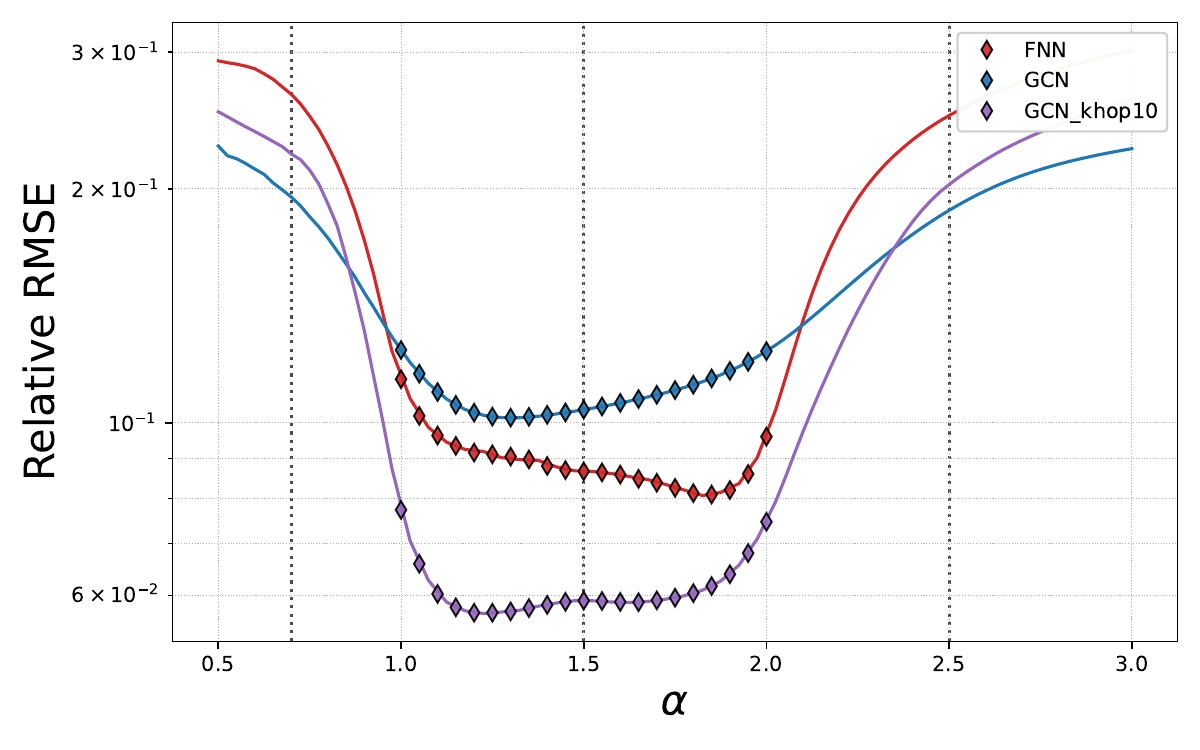}
        \caption{3 inputs}
        \label{fig:error_FNN&GNN_3input}
    \end{subfigure} 
   
    \begin{subfigure}{0.47\textwidth}
        \includegraphics[width=\linewidth]{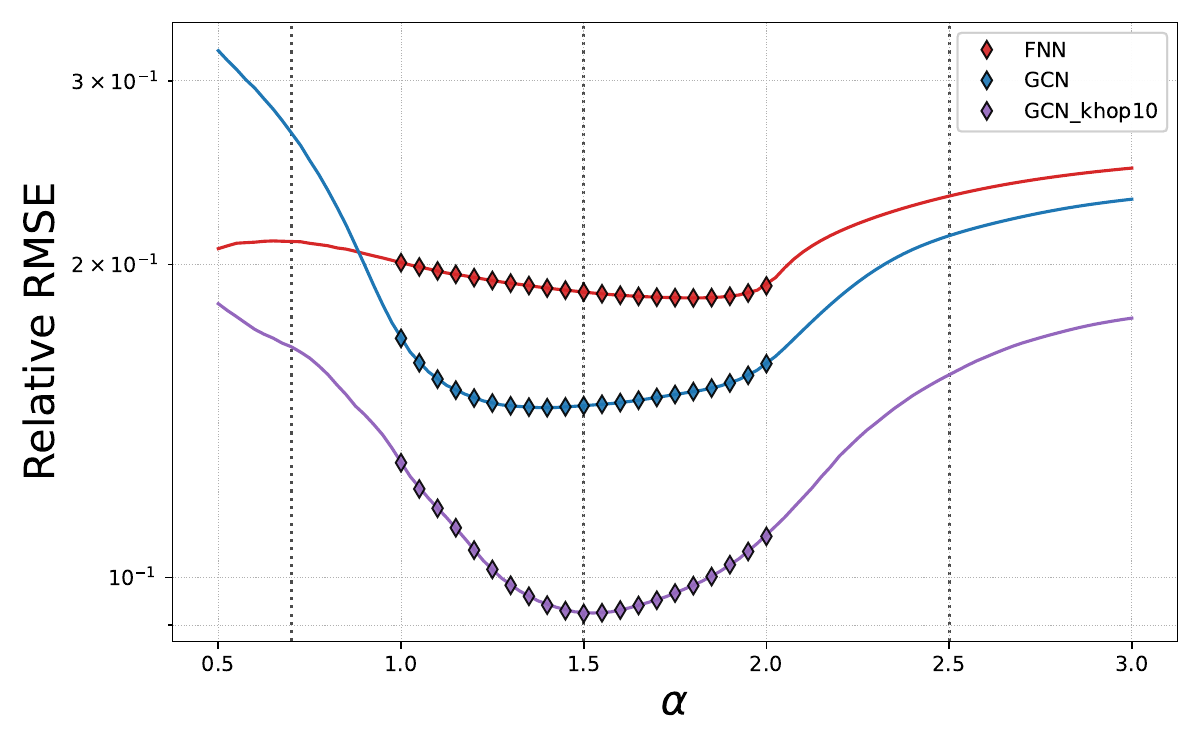}
        \caption{2 inputs}
        \label{fig:error_FNN&GNN_2input}
    \end{subfigure}
    \hfill
    \begin{subfigure}{0.48\textwidth}
        \includegraphics[width=\linewidth]{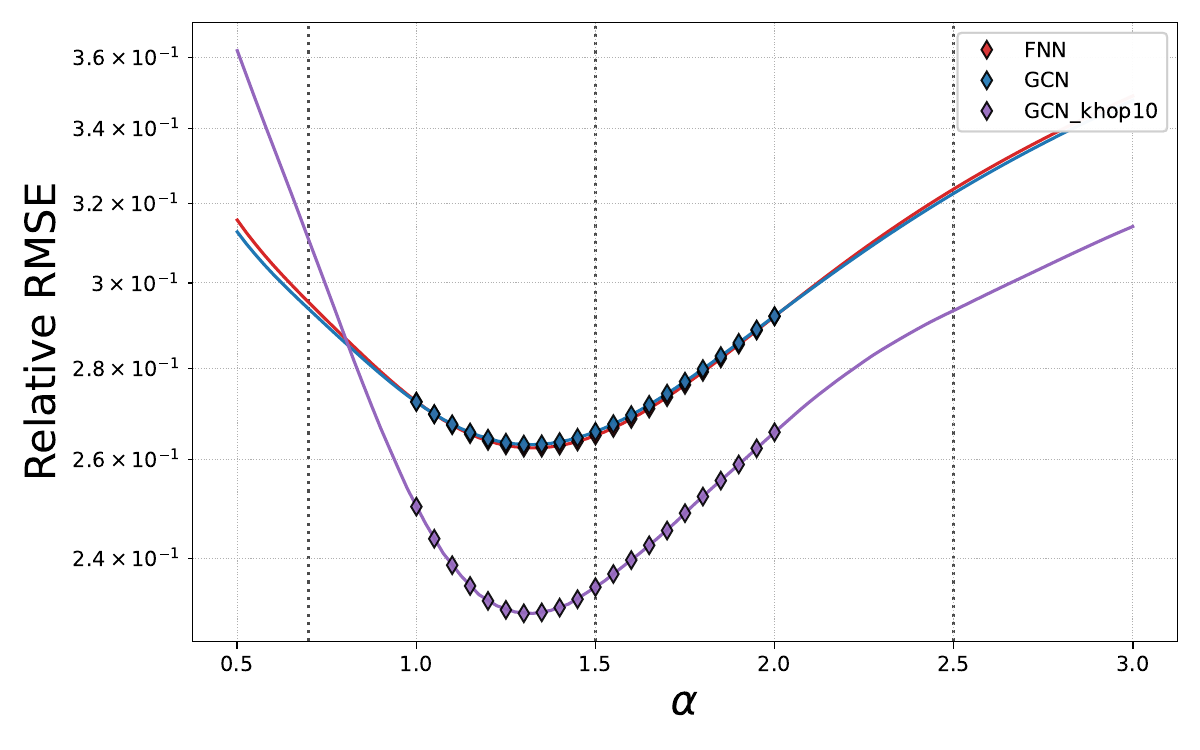}
        \caption{1 input}
        \label{fig:error_FNN&GNN_1input}
    \end{subfigure}
    \caption{RRMSE as a function of $\alpha$ for architectures with a
  reduced number of input features: (\textit{a}) 4 inputs; (\textit{b}) 3 inputs; (\textit{c}) 2 inputs; (\textit{d}) 1 input.}\label{fig:GCNk&FNN_architecture_comparison}
\end{figure}

Figure~\ref{fig:GCNk&FNN_architecture_comparison} shows the RRMSE as a function of $\alpha$ for \texttt{FNN}, \texttt{GCN} and \texttt{GCN\_khop}, for models with 4, 3, 2 and 1 input feature, panels \ref{fig:error_FNN&GNN_4input} to \ref{fig:error_FNN&GNN_1input} respectively, each with the same architecture, with the exception of the input layer.
As the number of input features decreases, the relative advantage of the graph based models over \texttt{FNN} increases. With four inputs, panel \ref{fig:error_FNN&GNN_4input}, the trend mirrors the five-input case of Section~\ref{sec:5inputs}, with \texttt{FNN} most accurate within the training range and \texttt{GCN\_khop} most accurate for the smallest $\alpha$ values. With three, two and one input, panels \ref{fig:error_FNN&GNN_3input} to \ref{fig:error_FNN&GNN_1input}, \texttt{FNN} becomes progressively less competitive, and \texttt{GCN\_khop} attains the lowest error across most of the range shown.
 
These trends are summarized in Tab.~\ref{tab:GNN_FNN_decreasing_input}, which also reports the validation loss and the number of training epochs for each configuration.
With four inputs, \texttt{FNN} achieves the lowest training error and \texttt{GCN\_khop} attains the lowest extrapolation errors. With three inputs, all errors increase considerably: \texttt{GCN\_khop} attains the lowest training error, while \texttt{FNN} becomes the least accurate model in both extrapolation regions. With two inputs, \texttt{GCN\_khop} is the most accurate model across all three metrics. With a single input, \texttt{FNN} and \texttt{GCN} perform almost identically, while \texttt{GCN\_khop} achieves the lowest error in the training range and in the upper extrapolation region.

The above analysis shows an important feature of graph-based models: when the information content of the inputs is low, as for the case with one or two inputs, non-local information exchanged through the graph structure partly compensates for the reduced input information. In contrast, with four or five inputs, the local features already carry most of the information needed to predict $\tau$, and \texttt{FNN} remains competitive within the training range. 

For complex flows, it is not always straightforward to identify all relevant input features. Our results suggest that graph-based models can compensate for reduced input information by exploiting the underlying spatial structure.

\begin{table}[ht!]
    \centering
    \caption{Model comparison for varying number of inputs.}
    \label{tab:GNN_FNN_decreasing_input}
    \sisetup{group-digits=false}
    \begin{tabular}{
        l
        c
        c
        S[scientific-notation=true]
        S[table-format=1.5]
        S[table-format=1.5]
        S[table-format=1.5]
    }
    \toprule
    \textbf{Model}
        & \textbf{\# inputs}
        & \textbf{Epochs}
        & {\textbf{Val. loss}}
        & {\textbf{Testing 1}}
        & {\textbf{Training}}
        & {\textbf{Testing 2}} \\
    \midrule

    FNN      & 4 & 1691 & 7.942e-06 & 0.05769 & 0.00765 & 0.02112 \\
    GCN      & 4 & 6000 & 1.553e-04 & 0.06074 & 0.03707 & 0.03853 \\
    GCN$\_{\mathrm{khop}}$
             & 4 & 6000 & 6.053e-05 & 0.05019 & 0.00981 & 0.02006 \\

    \midrule

    FNN      & 3 & 1078 & 1.037e-03 & 0.23515 & 0.08864 & 0.23269 \\
    GCN      & 3 & 6000 & 1.418e-03 & 0.18297 & 0.10852 & 0.18238 \\
    GCN$\_{\mathrm{khop}}$
             & 3 & 6000 & 6.709e-04 & 0.19400 & 0.06058 & 0.18772 \\

    \midrule

    FNN      & 2 & 534  & 4.602e-03 & 0.20778 & 0.18997 & 0.22903 \\
    GCN      & 2 & 6000 & 2.756e-03 & 0.25224 & 0.15047 & 0.20697 \\
    GCN$\_{\mathrm{khop}}$
             & 2 & 6000 & 1.967e-03 & 0.16063 & 0.10142 & 0.15209 \\

    \midrule

    FNN      & 1 & 263  & 9.494e-03 & 0.29296 & 0.27099 & 0.32259 \\
    GCN      & 1 & 6000 & 9.201e-03 & 0.29149 & 0.27155 & 0.32166 \\
    GCN$\_{\mathrm{khop}}$
             & 1 & 6000 & 7.705e-03 & 0.30378 & 0.24182 & 0.29239 \\

    \bottomrule
    \end{tabular}
\end{table}

\newpage
\subsection{Out-of-distribution configurations}\label{sec:ood}
 
\begin{figure}
    \centering
    \includegraphics[width=0.9\linewidth]{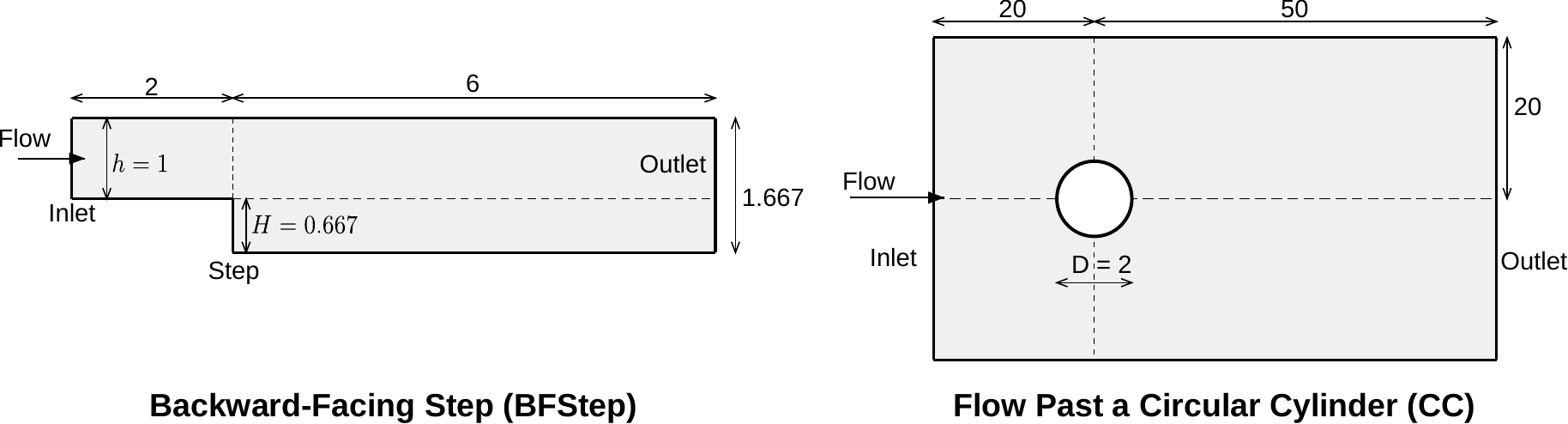}
    \caption{Domains of the backward-facing step (BFStep) and circular-cylinder (CC) cases used to investigate the generalization ability of the models.}
    \label{fig:OOD_domains}
\end{figure}

The generalization capability of the three models introduced in Section~\ref{sec:5inputs}, trained on the PH dataset at $\mathrm{Re} = 100$, with 5 input features, is assessed on two configurations not included in training: a backward-facing step (BFStep), at $\mathrm{Re} = 66.667$ based on the step height, and a steady flow past a circular cylinder (CC), at $\mathrm{Re} = 40$. The BFStep case can be interpreted as a limiting case of the PH geometry, with an effectively vertical slope, and shares the same flow-separation mechanism, while introducing an abrupt geometric discontinuity absent from the training set. The CC one, instead, introduces a fundamentally different external flow configuration. Figure~\ref{fig:OOD_domains} schematically represents the domains of the two investigated flows.

For each configuration, the corresponding table reports the RRMSE together with three further metrics computed on the raw predicted and reference $\tau$ fields: the percentage of mesh cells with a local relative error below $5\%$ and below $10\%$, denoted \texttt{pct\_below\_5pct} and \texttt{pct\_below\_10pct} respectively, and the relative bias, \texttt{bias\_rel}, computed as the mean signed error, $\tau^{\mathrm{pred}} - \tau^{\mathrm{sim}}$, normalized by the mean of $|\tau^{\mathrm{sim}}|$. A positive \texttt{bias\_rel} indicates systematic over-prediction, a negative one systematic under-prediction.

\subsubsection{Backward Facing Step}\label{sec:bfs}
 
\begin{figure}[h!]
    \centering
    \begin{subfigure}{0.48\textwidth}
        \centering
        \includegraphics[width=\textwidth]{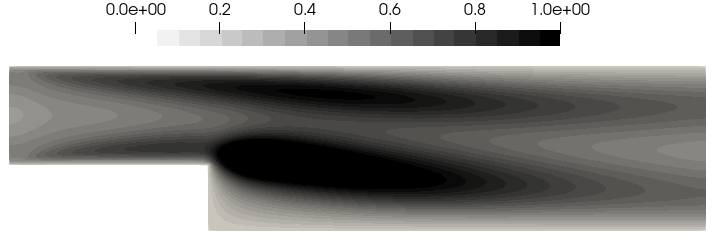}
        \caption{$\tau$}
    \end{subfigure}
    \hfill
    \begin{subfigure}{0.48\textwidth}
        \includegraphics[width=\linewidth]{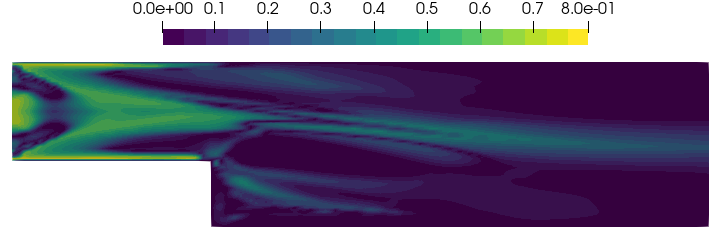}
        \caption{FNN}
    \end{subfigure} 
   
    \begin{subfigure}{0.48\textwidth}
        \includegraphics[width=\linewidth]{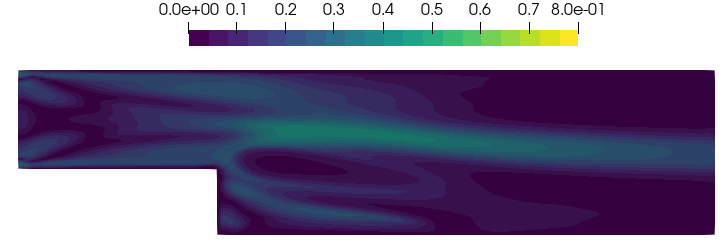}
        \caption{GCN}
    \end{subfigure}
    \hfill
    \begin{subfigure}{0.48\textwidth}
        \includegraphics[width=\linewidth]{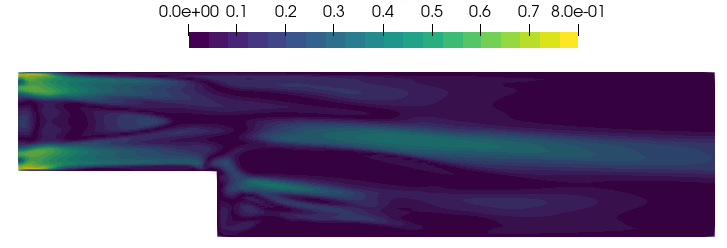}
        \caption{GCN\_khop}
    \end{subfigure}

    \caption{Absolute error of the scalar field $\tau$, clipped to $[0,1]$, on the BFStep configuration: (\textit{a}) reference field $\tau$; (\textit{b}) FNN; (\textit{c}) GCN; (\textit{d}) GCN\_khop ($k$-hop = 10). All models use 10551 trainable parameters.} \label{fig:abs_errprediction_tau_BFS}
\end{figure}

Figure~\ref{fig:abs_errprediction_tau_BFS} shows the absolute error of the predicted $\tau$ field for the BFStep configuration. Because all models have a sigmoid activation function on the output layer, the reference $\tau$ field has been set equal to 1 whenever it attained higher values. In this manner, we can isolate the error associated to models prediction. The largest errors for all three models are concentrated downstream of the step, spanning the recirculation region and the reattachment zone, where the flow structure differs most from the PH training set. Both \texttt{GCN} and \texttt{GCN\_khop} exhibit lower errors than the \texttt{FNN} in this region. A second, distinct error region appears upstream of the step, in the inlet section of the channel: here \texttt{FNN} shows a wide band of error spanning the full channel height, \texttt{GCN\_khop} shows a smaller but still visible error patch, while \texttt{GCN} remains close to the reference solution throughout this region.

Tab.~\ref{tab:BFS_summary} reports the corresponding quantitative metrics. \texttt{GCN} attains the lowest RRMSE, equal to 0.2707, followed by \texttt{GCN\_khop} at 0.2989 and \texttt{FNN} at 0.4056, confirming the qualitative trend. Regarding the relative bias,  \texttt{GCN} is the most biased model, with a \texttt{bias\_rel} of -0.2048, indicating a systematic under prediction of $\tau$ across the domain; \texttt{GCN\_khop} is less biased, at -0.0703; \texttt{FNN}, despite the largest overall RRMSE, has a \texttt{bias\_rel} close to zero, at 0.0245. This indicates that \texttt{FNN}'s over- and under-predictions are more evenly split across the domain, rather than skewed consistently in one direction as for \texttt{GCN} and \texttt{GCN\_khop}. This is also reflected in the percentage of cells within $5\%$ and $10\%$ relative error: \texttt{FNN} attains the highest values, $20.87\%$ and $38.02\%$ respectively, against $17.43\%$ and $31.18\%$ for \texttt{GCN} and $16.90\%$ and $32.29\%$ for \texttt{GCN\_khop}. \texttt{FNN} therefore predicts a larger fraction of individual cells accurately, while \texttt{GCN} achieves a lower global error together with a more systematic, one-directional bias. This property, unlike unbiased errors, could in principle support the definition of a targeted correction strategy.

\begin{table}[ht!]
\caption{Summary of BFStep results.}
\label{tab:BFS_summary}
\begin{tabular}{lcccc}
\toprule
 \textbf{Model} &  RRMSE & \textbf{pct\_below\_5pct} & \textbf{pct\_below\_10pct} & \textbf{bias\_rel} \\
\midrule
 FNN & 0.4056 & 20.8667 & 38.0222 & 0.0245 \\
 GCN & 0.2707 & 17.4333 & 31.1778 & -0.2048 \\
 GCN\_khop & 0.2989 & 16.900 & 32.2889 & -0.0703 \\
\bottomrule
\end{tabular}
\end{table}

\subsubsection{Circular Cylinder}\label{sec:cc}
 
\begin{figure}[h!]
    \centering
    \begin{subfigure}{0.48\textwidth}
        \centering
        \includegraphics[width=\textwidth]{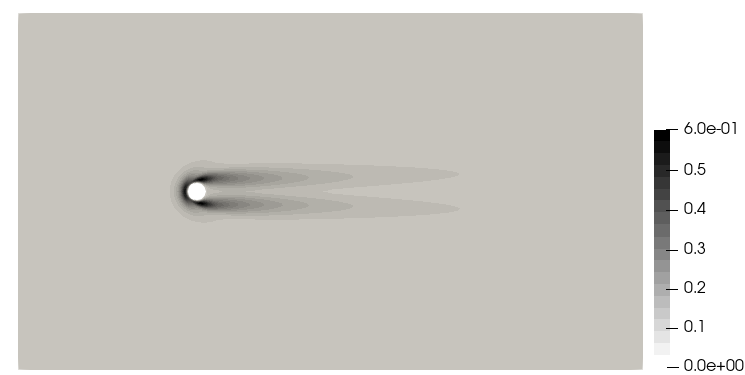}
        \caption{$\tau$}
    \end{subfigure}
    \hfill
    \begin{subfigure}{0.48\textwidth}
        \includegraphics[width=\linewidth]{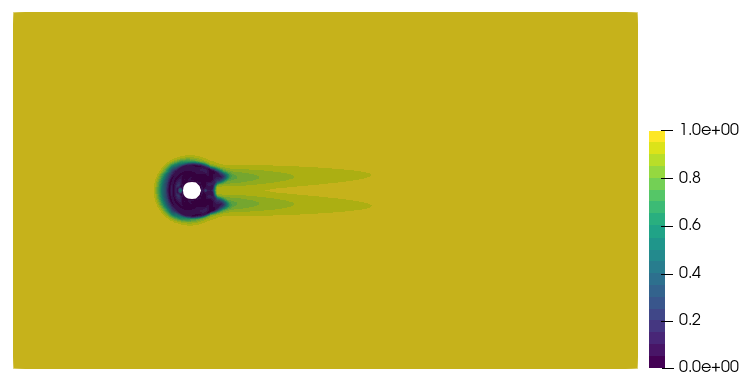}
        \caption{FNN}
    \end{subfigure} 
   
    \begin{subfigure}{0.48\textwidth}
        \includegraphics[width=\linewidth]{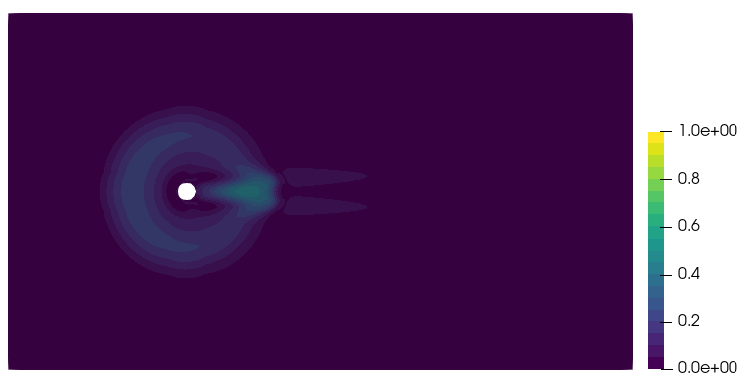}
        \caption{GCN}
    \end{subfigure}
    \hfill
    \begin{subfigure}{0.48\textwidth}
        \includegraphics[width=\linewidth]{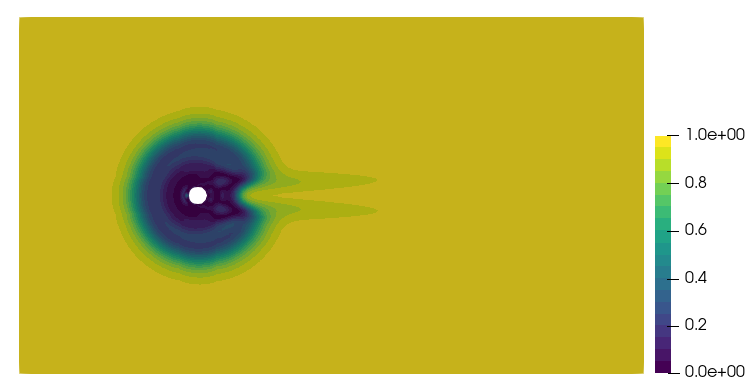}
        \caption{GCN\_khop}
    \end{subfigure}

    \caption{Absolute error of the scalar field $\tau$ on the Circular Cylinder configuration: (\textit{a}) reference field $\tau$; (\textit{b}) FNN; (\textit{c}) GCN; (\textit{d}) GCN\_khop ($k$-hop = 10). All models use 10551 trainable parameters.} \label{fig:abs_errprediction_tau_CC}
\end{figure}

Figure~\ref{fig:abs_errprediction_tau_CC} shows the comparison for the CC configuration, an external flow configuration absent from the training set. Thus, it is a markedly more challenging case for the models. \texttt{GCN} error remains low across most of the domain, with a ring-shaped region of elevated error immediately surrounding the cylinder. \texttt{FNN} and \texttt{GCN\_khop} show the opposite pattern: a widespread, near-uniform high error across most of the domain, interrupted only by a localized low error region around the cylinder and its wake, where the models, near constant, over prediction happens to match the elevated reference field there. The \texttt{GCN\_khop}'s low-error region narrows more rapidly with distance from the cylinder than \texttt{GCN}'s, consistent with the reduced receptive field of the sampled model.

Two further tests were carried out to investigate the origin of \texttt{GCN}'s ring. Increasing the depth of \texttt{GCN\_khop} to $k= 20$ and $k = 50$ did not change the ring's spatial extent, ruling out the sampling radius as its cause. Removing the wall distance feature, in addition to the mixing length already excluded in the four-input configuration (Section~\ref{sec:decreasing_inputs}), removed the ring but increased the overall error, consistent with the loss of two informative inputs. This suggests the wall distance feature is distributed differently on the CC geometry than on the periodic hills used for training, and is the most likely origin of the ring.

Tab.~\ref{tab:CC_summary} reports the quantitative metrics. \texttt{GCN} is markedly more accurate than the other two models, with a RRMSE of 0.9347, against 7.5739 for \texttt{GCN\_khop} and 8.5226 for \texttt{FNN}. For \texttt{FNN} and \texttt{GCN\_khop}, the RRMSE exceeds unity, meaning the prediction error is larger than the norm of the reference field itself: their predictions are effectively unusable on this configuration. This behavior is explained by the relative bias: \texttt{FNN} and \texttt{GCN\_khop} show large positive \texttt{bias\_rel} values, indicating a strong systematic over prediction of $\tau$, consistent with the widespread high error background described above. \texttt{GCN}, in contrast, has a \texttt{bias\_rel} still indicating some over-prediction but far more contained.
The percentage of cells within $5\%$ and $10\%$ relative error is low for all three models, between $1.66\%$ and $2.31\%$ and between $3.72\%$ and $4.76\%$ respectively, reflecting the difficulty of this out-of-distribution configuration even for the best-performing model.

\begin{table}[ht!]
\caption{Summary of CC results.}
\label{tab:CC_summary}
\begin{tabular}{lcccc}
\toprule
\textbf{Model} &  RRMSE & \textbf{pct\_below\_5pct} & \textbf{pct\_below\_10pct} & \textbf{bias\_rel} \\
\midrule
 FNN & 8.5226 & 2.3125 & 4.1429 & 19.4028 \\
 GCN & 0.9347 & 1.6607 & 4.7589 & 1.1389 \\
 GCN\_khop & 7.5739 & 1.8571 & 3.7232 & 16.2078 \\
\bottomrule
\end{tabular}
\end{table}

Taken together, the quantitative and qualitative results discussed in this section indicate that non-local information exchanged through the graph structure improves the accuracy of the predicted $\tau$ field, both within the training distribution, if the input information is not sufficient, and, more markedly, outside it. Sampling the graph structure during training, as in \texttt{GCN\_khop}, further improves accuracy over \texttt{GCN} within and close to the training range, although this advantage narrows as $\alpha$ moves further from the training interval, where the accuracy of \texttt{GCN\_khop} approaches that of \texttt{GCN}. This effect is most pronounced on the CC configuration, where \texttt{GCN\_khop} and \texttt{FNN} attain a RRMSE above unity while \texttt{GCN} remains below one: access to information from the full mesh, rather than from a locally sampled neighborhood, becomes critical when extrapolating to flow configurations markedly different from the training distribution.

\vspace{1cm}
\section{Results: RANS closure challenge}\label{sec:results_rans}
\subsection{Models hyperparameters selection and training strategies}\label{sec:rans_hyperparams}

The methodology developed for the Transport benchmark (Section~\ref{sec: hyperparameters and training}) is applied unchanged to the RANS closure problem~\cite{closurechallenge}. Specifically, the same architecture is kept for the three models \texttt{FNN}, \texttt{GCN} and \texttt{GCN\_khop} with $k=10$, each with 10501 trainable parameters, and the same training configuration of Table~\ref{tab:training_config} is used. As already mentioned in Section~\ref{sec:rans}, we do not include the mixing length input in RANS experiments. Hence, the number of parameters is reduced compared to the Transport phenomena flow. The training simulations are 21, making the number of subgraphs used to train the \texttt{GCN\_khop} model equal to 8400.

For the sake of conciseness, we do not report a systematic analysis of the sampling depth $k$, postponed for the interested reader in Appendix \ref{app:gcn_rans_khop_comparison}, nor ablation studies as in Section~\ref{sec:results_transport}. Generally speaking, we observed similar qualitative behaviors as for the Transport phenomenon case.

The only noteworthy differences observed for the \texttt{GCN} training concern the batch size, decreased from 5 to 2 graphs, and the scheduler patience for the learning-rate updates, increased from 40 to 400 epochs. The smaller batch size results in a larger number of parameter updates for a given number of training graphs, while the increased patience compensates for the smaller number of optimization steps performed by \texttt{GCN} per epoch on this dataset. These modifications were found to be necessary to improve the training performance of the \texttt{GCN}. To illustrate this effect, we refer the reader to Appendix~\ref{app:gcn_rans_training}, where the results obtained with scheduler patience values of 40 and 400 epochs are compared.

We point out that, despite the larger patience, \texttt{GCN} undergoes fewer optimization steps, and hence fewer weight updates, than both \texttt{FNN} and \texttt{GCN\_khop}, as quantified in Appendix~\ref{app:gcn_rans_4inputs_training}. Therefore, the comparison that will be presented in this Section is not biased in favour of \texttt{GCN} by the increased scheduler patience.

\subsection{Comparative Analysis: quantitative and qualitative results}\label{sec:rans_modelscomparison}

\begin{figure}[ht!]
    \centering
    \begin{subfigure}{0.9\textwidth}
        \centering
        \includegraphics[width=\textwidth]{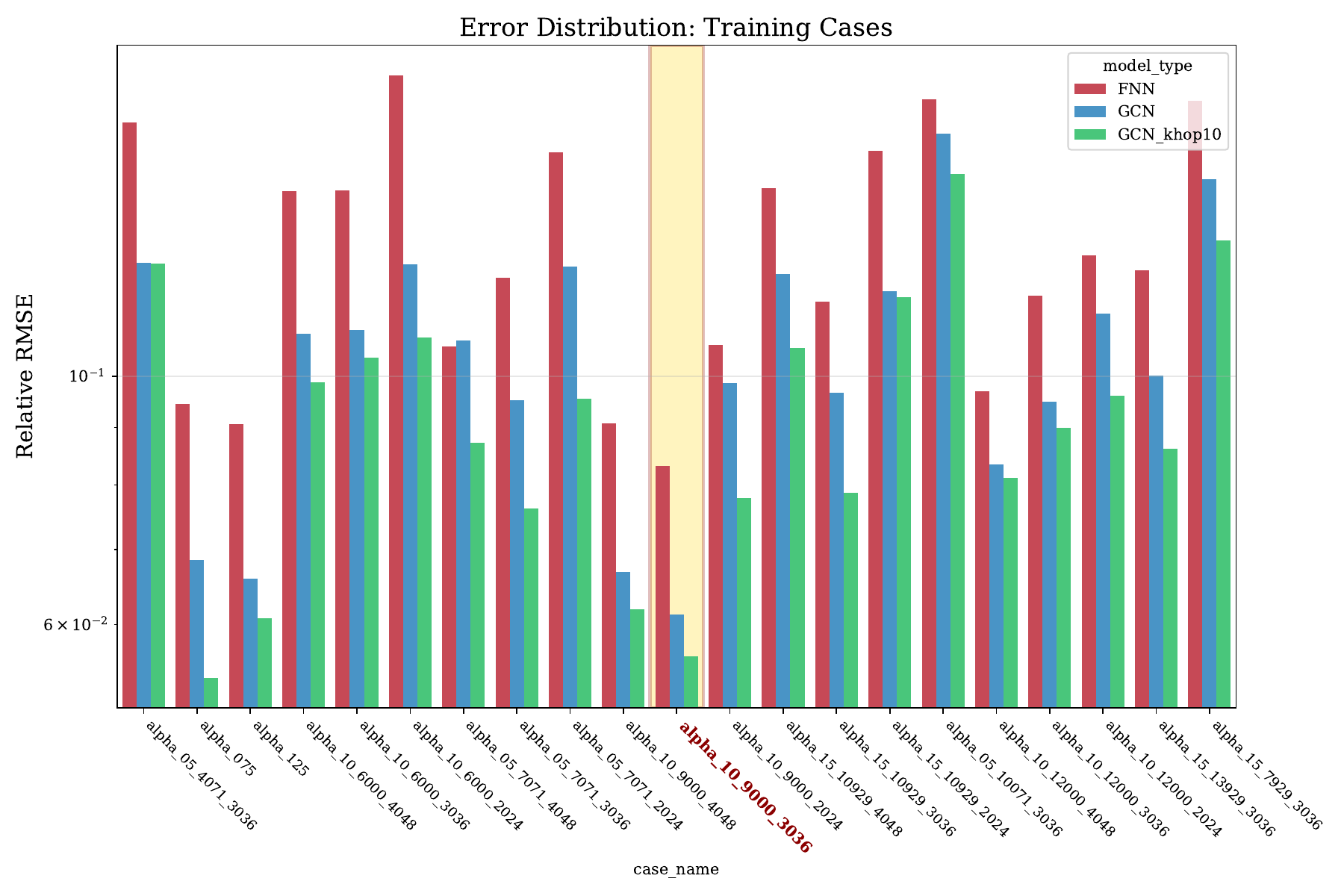}
        \caption{Training}
    \end{subfigure}
    \hfill
    \begin{subfigure}{0.9\textwidth}
        \includegraphics[width=\linewidth]{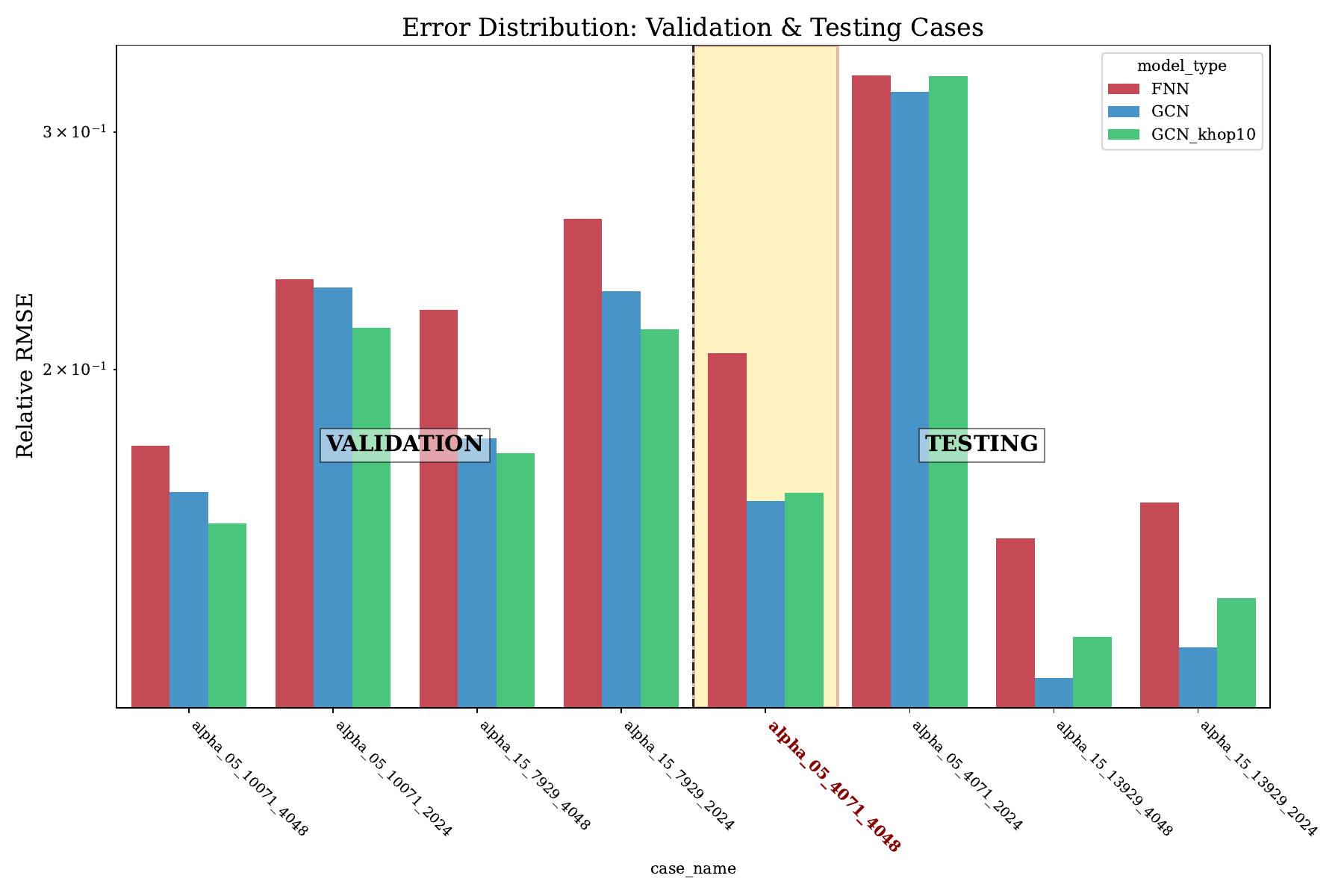}
        \caption{Validation and Testing}
    \end{subfigure} 
   
    \caption{Comparison between the \texttt{FNN}, the \texttt{GCN} and \texttt{GCN\_khop}, with $k=10$, models: A) RRMSE with respect to the cases used for training; B) RRMSE with respect to the cases used for validation and testing.}\label{fig:GCN_FNN_Khop_RANS_error}
\end{figure}

Fig.~\ref{fig:GCN_FNN_Khop_RANS_error} shows, for each configuration, the RRMSE of \texttt{FNN}, \texttt{GCN} and \texttt{GCN\_khop} for the prediction of $\nu_t$, on the training, validation and testing sets. The three models attain errors of the same order of magnitude on every configuration, but with systematic trends. \texttt{GCN\_khop} attains the lowest error on nearly every training configuration and on all four validation configurations, while \texttt{GCN} attains the lowest error on all four testing configurations, by a small margin on two of them. \texttt{FNN} is the least accurate of the three models, except for a testing configuration where its error equals the \texttt{GCN\_khop} one.

These trends are summarized in Tab.~\ref{tab:RANS_GNN_FNN_10501tp_4input_check}, which also reports the validation loss and the number of training epochs. \texttt{GCN\_khop} converges in the fewest epochs, equal to 408, and attains the lowest validation loss, equal to $7.206\times10^{-6}$. \texttt{FNN} converges in 811 epochs, with a validation loss equal to $9.51\times10^{-6}$. \texttt{GCN}, at the increased scheduler patience, uses the full 6000 epoch budget and attains a validation loss of $8.079\times10^{-6}$, between the other two models, in agreement with the validation errors of Figure \ref{fig:GCN_FNN_Khop_RANS_error}.

The relative errors confirm that the three models are close to one another. On the training set, \texttt{GCN\_khop} attains the lowest error, followed by \texttt{GCN} and \texttt{FNN}. The same ordering holds on the validation set. On the testing set, \texttt{GCN} attains the lowest error, followed by \texttt{GCN\_khop} and \texttt{FNN}. The two graph-based models differ from each other by at most 14\%, and by less than 6\% on validation and testing, while \texttt{FNN} is between 17\% and 41\% higher than the best model on each split. Contrary to the Transport case, in the RANS scenario the \texttt{FNN} model underperforms graph-based models both on validation/testing and on training configurations. In addition, consistently with the analysis carried out in the Transport flow section, the \texttt{GCN\_khop} model outperforms the \texttt{GCN} on training and validations configurations, while the latter seems to generalize better on unseen configurations, as testified by a slightly smaller testing error.

\begin{table}[ht!]
    \centering
    \caption{Model comparison for the RANS flow}
    
    \begin{tabular}{lccccc}
    \hline
    \textbf{Model} & \textbf{Epochs} & \textbf{Val\_loss} & \textbf{err\_validation} &\textbf{err\_training} & \textbf{err\_testing}  \\
    \hline
    FNN & 811 & 9.51e-06 & 0.2223 & 0.1307 & 0.2114 \\
    GCN & 6000 & 8.079e-06 & 0.1997 & 0.1051 & 0.1809 \\
    GCN\_khop & 408 & 7.206e-06 & 0.1891 & 0.0927 & 0.1886 \\
    \hline 
    \end{tabular}
    \label{tab:RANS_GNN_FNN_10501tp_4input_check}
\end{table}

\begin{figure}[h!]
    \centering
    \begin{subfigure}{0.48\textwidth}
        \centering
        \includegraphics[width=\textwidth]{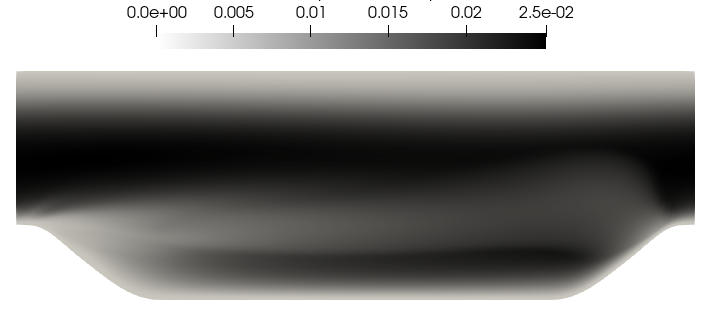}
        \caption{$\nu_t$}
    \end{subfigure}
    \hfill
    \begin{subfigure}{0.48\textwidth}
        \includegraphics[width=\linewidth]{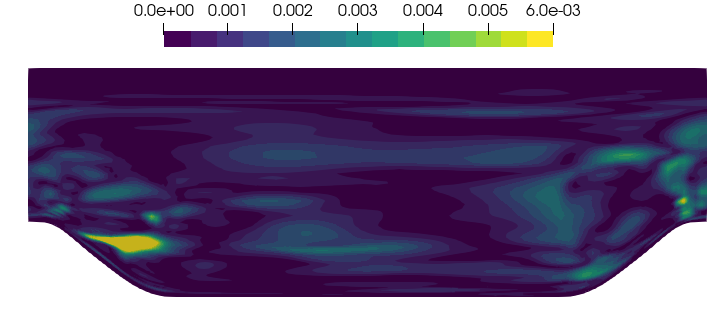}
        \caption{FNN}
    \end{subfigure} 
   
    \begin{subfigure}{0.48\textwidth}
        \includegraphics[width=\linewidth]{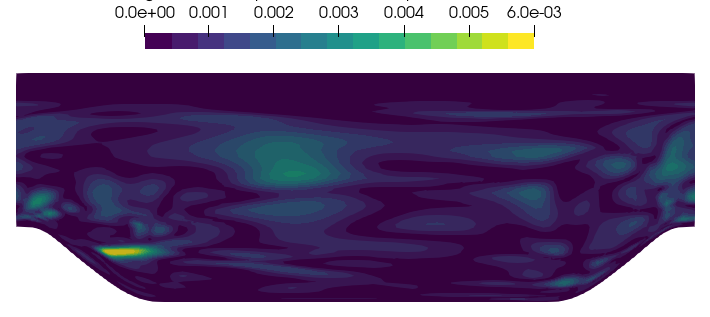}
        \caption{GCN}
    \end{subfigure}
    \hfill
    \begin{subfigure}{0.48\textwidth}
        \includegraphics[width=\linewidth]{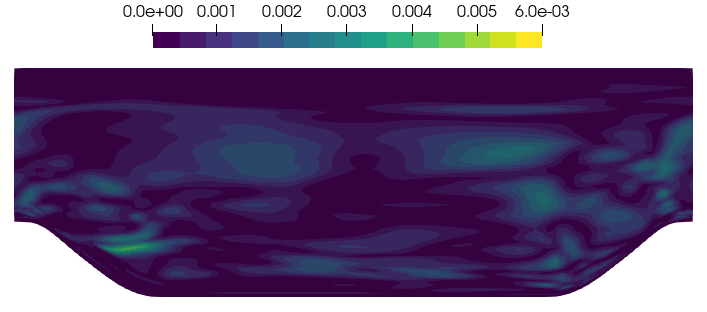}
        \caption{GCN\_khop}
    \end{subfigure}

    \caption{Absolute error of the scalar field $\nu_{t}$ on the PH configuration with $\alpha = 1.0$, $L_x/H = 9.000$ and $L_y/H = 3.036$: (\textit{a}) reference field $\nu_{t}$; (\textit{b}) FNN; (\textit{c}) GCN; (\textit{d}) GCN\_khop ($k = 10$).} \label{fig:abs_err_nut_PH_train}
\end{figure}

\begin{figure}[h!]
    \centering
    \begin{subfigure}{0.44\textwidth}
        \centering
        \includegraphics[width=\textwidth]{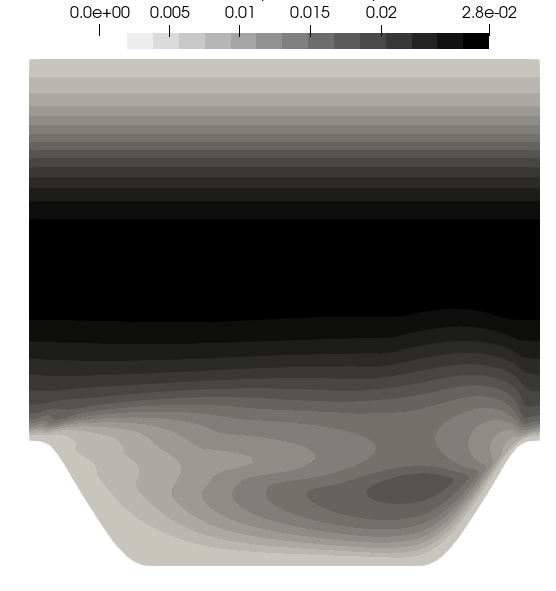}
        \caption{$\nu_t$}
    \end{subfigure}
    \hfill
    \begin{subfigure}{0.48\textwidth}
        \includegraphics[width=\linewidth]{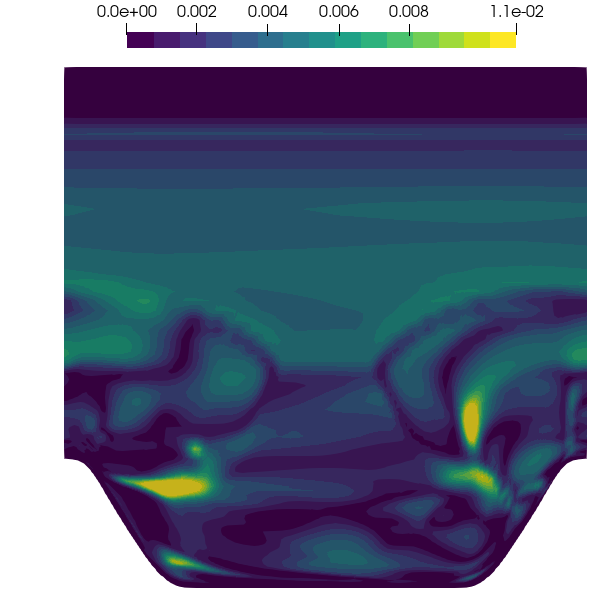}
        \caption{FNN}
    \end{subfigure} 
   
    \begin{subfigure}{0.48\textwidth}
        \includegraphics[width=\linewidth]{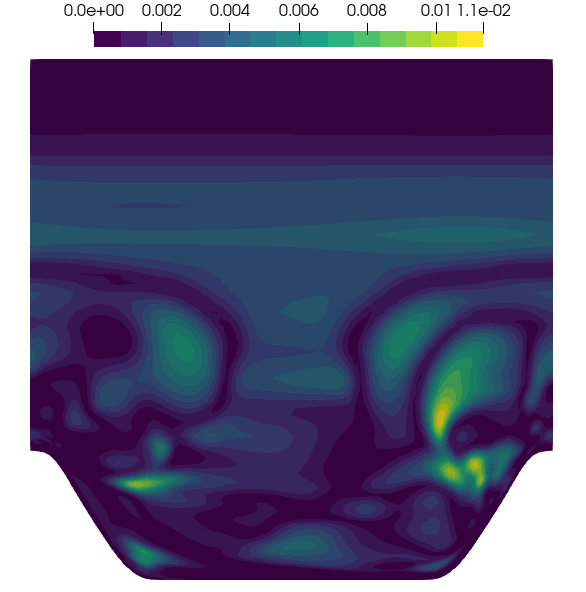}
        \caption{GCN}
    \end{subfigure}
    \hfill
    \begin{subfigure}{0.47\textwidth}
        \includegraphics[width=\linewidth]{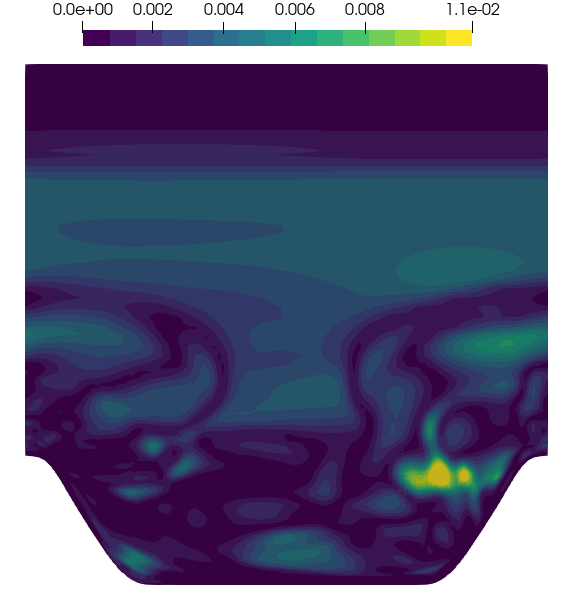}
        \caption{GCN\_khop}
    \end{subfigure}

    \caption{Absolute error of the scalar field $\nu_{t}$ on the PH configuration with $\alpha = 0.5$, $L_x/H = 4.071$ and $L_y/H = 4.048$: (\textit{a}) reference field $\nu_{t}$; (\textit{b}) FNN; (\textit{c}) GCN; (\textit{d}) GCN\_khop ($k= 10$).} \label{fig:abs_err_nut_PH_test}
\end{figure}

Figures~\ref{fig:abs_err_nut_PH_train} and~\ref{fig:abs_err_nut_PH_test} show the absolute error of the predicted $\nu_t$ field on two representative PHs configurations: a training configuration with $\alpha = 1.0$, and a testing configuration with $\alpha = 0.5$. 

For the training configuration (Fig.~\ref{fig:abs_err_nut_PH_train}), the error is low over most of the domain for all three models, and the largest values are concentrated in a few localized regions. The highest error is located at a sharp spot on the left slope of the domain, at the bottom left, immediately downstream of the hill crest. The spot is clearly visible for \texttt{FNN}, present but narrower for \texttt{GCN}, and much weaker for \texttt{GCN\_khop}, which shows only a faint streak along the same slope. Elsewhere, patches of moderate error appear in the central part of the domain and near the right slope for all models. The \texttt{FNN} error map is the most irregular, with a larger number of patches of intermediate error than the two graph-based models, consistently with its higher training error (Tab.~\ref{tab:RANS_GNN_FNN_10501tp_4input_check}).

For the testing configuration (Fig.~\ref{fig:abs_err_nut_PH_test}), the errors are larger and concentrated in the lower half of the domain, in the valley and in the shear layer above it, while the upper channel is predicted with low error by all models. A band of moderate error, just below the upper region, is visible for all three models. Within the valley, the models struggle to predict $\nu_t$ in elongated, curved structures. \texttt{FNN} shows an intense error streak on the left side of the valley and a large arc-shaped structure on the right side, with an elongated peak of error. \texttt{GCN} reproduces the right-hand structure with comparable intensity, while the left-hand spot is considerably weaker. \texttt{GCN\_khop} shows a weaker right-hand arc, but concentrates its largest error in a compact spot in the lower central-right part of the valley, much less pronounced in the other two models. The three models therefore struggle in the same part of the domain, the valley and the regions around the slopes, with differences in the location and intensity of the peaks rather than in the overall error level, consistently with the small differences of Tab.~\ref{tab:RANS_GNN_FNN_10501tp_4input_check}.

We do not report results of the BFStep flow, because the obtained errors for all models are excessively high. We think that using a tailored model for RANS problems will make such analysis meaningful for future works.
 
To conclude this section, we point out that the relative RRMSE values observed in this RANS test case are significantly higher than those obtained for the Transport case. This difference is an expected consequence of the methodological choices adopted in the present framework. Indeed, although the RANS flow is intrinsically more challenging than the Transport case, both because of its turbulent nature and its increased geometrical complexity, we deliberately retain the same model architectures, in terms of trainable parameters, and the same general strategy for input selection. Specifically, the inputs considered for the RANS case are not tailored to this flow and, in addition, one input has been discarded to avoid an algebraic dependence of the output on the inputs. These choices are motivated, on the one hand, by the need to maintain a clear and consistent comparison between the two cases and, on the other hand, by the objective of providing a proof-of-concept demonstrating whether the same learning framework can be extended from a relatively simple laminar setting to a substantially more challenging turbulent flow problem.

Unlike on the Transport benchmark, where the advantage of the graph-based models became clear as the number of input features was reduced (Section~\ref{sec:decreasing_inputs}), on the RANS benchmark the three models attain comparable accuracy, with a small advantage of the graph-based ones over \texttt{FNN}. The non-local character of turbulent flows provides a physical motivation for exploiting information beyond the immediate local flow features, and the ability of graph-based models to incorporate information from the surrounding computational mesh can partially compensate for the limited local information provided as input. The RANS benchmark therefore provides a first indication of the potential benefit of exploiting the spatial structure of the computational mesh. A more tailored choice of the input features, or the use of larger models , should be investigated in future works to isolate the benefits of using non-local models to handle non-locality of turbulence only.

\vspace{2cm}

\newpage
\section{Conclusions and perspectives} \label{sec:conclu}

In this work, we compared FNNs and GCNs for the ML-based RANS closure problem on two periodic hills flows. The FNN architecture is a paradigmatic one for local models, while the GCN architecture exploits the underlying graph structure of the computational mesh to provide non-local effects. Our findings show that the FNN provided accurate predictions for the laminar case within the training regime and with input features that carry sufficient information. However, its performance degraded if one of the two conditions was not satisfied. In contrast, the GCN generalized better to unseen geometries and compensated for missing input information. On the turbulent case, the GCN also outperformed the FNN. These results indicate that non-local information improves robustness, and that GCNs are a promising tool for ML-based RANS modeling.
Since each simulation yields a single training graph, the number of training samples for the GCN is limited; we tested a data augmentation strategy based on the $k$-hop sub-sampling strategy. Specifically, from each simulation graph we extract a fixed number of subgraphs by randomly selecting the seed node and extracting neighbors up to order $k$. In the laminar flow it proved to be an effective strategy to increase GCN's accuracy in the training regime both with and without sufficiently informative input features. In the turbulent case, using this strategy did not significantly changed the performances, with errors slightly smaller for training configurations and slightly higher for testing configurations. This preliminary analysis suggests that such a procedure could be promising for more complex, 3D simulations with bigger computational meshes for which it is computationally infeasible to exploit the full associated graphs. 

Overall, this work shows the potential of graph-based ML models, namely GCNs, for RANS turbulence modeling. Since the turbulent case was treated as a proof-of-concept, with input features not tailored to RANS flows, a natural next step is to assess the accuracy of the method when features and architecture are designed specifically for turbulence modeling, and to test the resulting closures within the RANS solver. A further direction concerns the message-passing scheme: in the present GCN, neighbors contribute through the mesh connectivity only, so the model cannot distinguish a neighbor upstream from one downstream, or a near one from a far one, except through the input features. Message passing that uses edge features, such as relative distances and directions between cells, or anisotropic aggregation schemes, would allow the model to exploit this geometric information, and could be combined with architectures that embed frame invariance when the target is a tensor or a vector. Finally, more sophisticated strategies to cope with the scarcity of training graphs should be investigated: for highly turbulent flows, only a few DNS or LES simulations are available, which makes GCN training a non-trivial task.

\section*{Acknowledgements}
D.O. and G.R. are members of the Gruppo Nazionale Calcolo Scientifico-Istituto Nazionale di Alta Matematica (GNCS-INdAM).

\bibliographystyle{plain}
\bibliography{biblio}
\newpage
\appendix

\appendix
\section{Investigating the input set and the training behavior of GCNs on the RANS dataset}
\label{app:gcn_rans_training}
\label{app:gcn_rans_4inputs_training}

This appendix documents the choice of the 4 input configuration adopted for the RANS benchmark, and the training behavior of the models in that configuration. The first subsection quantifies the gap between the models with 5 inputs, which include the mixing length, and with 4 inputs. The remaining subsections concern the 4 input configuration: the optimization budget of the three models, the effect of the scheduler patience on \texttt{GCN}, and the OOD test. The batch size used for each model is 2 for \texttt{GCN}, 32 for \texttt{GCN\_khop} and 1024 for \texttt{FNN}, and is kept fixed throughout.

\subsection{Effect of the mixing length.}
The 5 input configuration of the RANS benchmark includes the mixing length $\ell_m$ among the input features, as in the transport benchmark. Tab.~\ref{tab:app_mixing_length} compares the three models trained with and without this feature, with architectures that differ only in the input layer.

\begin{table}[htbp]
  \centering
  \caption{Effect of the mixing length on the RANS results: models with 5 inputs (mixing length included, 10551 trainable parameters) and with 4 inputs (mixing length excluded, 10501 trainable parameters). \texttt{GCN} trained at scheduler patience 400.}
  \label{tab:app_mixing_length}
  \begin{tabular}{llcccc}
    \toprule
    Model & Inputs & Val.\ loss & err.\ training & err.\ validation & err.\ testing \\
    \midrule
    \texttt{FNN}       & 5 & $8.272\times10^{-7}$ & 0.0369 & 0.0550 & 0.0679 \\
    \texttt{FNN}       & 4 & $9.51\times10^{-6}$  & 0.1307 & 0.2223 & 0.2114 \\
    \midrule
    \texttt{GCN}       & 5 & $7.90\times10^{-7}$  & 0.0376 & 0.0590 & 0.0634 \\
    \texttt{GCN}       & 4 & $8.079\times10^{-6}$ & 0.1051 & 0.1997 & 0.1809 \\
    \midrule
    \texttt{GCN\_khop} & 5 & $5.980\times10^{-7}$ & 0.0243 & 0.0448 & 0.0624 \\
    \texttt{GCN\_khop} & 4 & $7.206\times10^{-6}$ & 0.0927 & 0.1891 & 0.1886 \\
    \bottomrule
  \end{tabular}
\end{table}

With five inputs, the three models reach nearly identical testing errors, between 0.0624 and 0.0679, within 9\% of one another, and \texttt{FNN}, a purely local model that cannot exploit the graph structure, attains an accuracy comparable to that of the graph-based models. Removing the mixing length increases the error of every model by a factor between 2.8 and 4.2 across the three splits, and the validation loss by about one order of magnitude.

This behavior follows from the closed-form relation between $\ell_m$, $\nu_t$ and $\|\boldsymbol{S}\|$ discussed in Section~\ref{sec:rans}: since $\|\boldsymbol{S}\|$ is also an input feature, a model receiving $\ell_m$ can recover the target pointwise from two of its own inputs. The low errors obtained with five inputs therefore do not reflect an ability to learn the closure from the flow features, and they hide any difference between local and non-local models. Retaining this feature would have made the comparison trivial; it was therefore excluded from the main analysis, so that the models must infer $\nu_t$ from the remaining features.

\subsection{Optimization budget across models.} The three models perform a markedly different number of weights updates even after the patience adjustment. With 15600 cells per configuration and 21 training configurations (327\,600 nodes total), \texttt{FNN} performs $\lceil 327\,600/1024 \rceil = 320$ updates per epoch and converges in 811 epochs, for 259\,520 updates in total. \texttt{GCN\_khop} performs approximately 263 updates per epoch on its 8400 sampled subgraphs and converges in 408 epochs, for approximately 107\,100 updates in total. \texttt{GCN}, even at patience 400, performs only 11 updates per epoch and uses the full 6000-epoch budget, for 66\,000 updates in total, fewer than either of the other two models.

The same asymmetry appears in the effective tolerance of the early-stopping criterion, expressed in updates rather than epochs. Training halts after twice the scheduler patience without improvement in the validation loss (Sec.~\ref{sec: hyperparameters and training}): 80 epochs at the default patience of 40, and 800 epochs for \texttt{GCN} at the increased patience of 400. In update terms, this corresponds to 25\,600 updates of tolerance for \texttt{FNN} and 21\,000 for \texttt{GCN\_khop} at their default patience, against only 8800 for \texttt{GCN} even at the increased patience of 400, less than half of what the other two models tolerate by default. The patience-400 result for \texttt{GCN} is therefore not obtained by granting it a more generous stopping criterion, in update terms, than the other two models: \texttt{GCN} remains both the least-trained and the least tolerant of the three by this measure, and the comparison of Sec.~\ref{sec:rans_modelscomparison} is not biased in its favor by this choice. At patience 400 it attains errors comparable to those of the other two models (Tab.~\ref{tab:app_patience_4input}) with the smallest number of updates in total.

\subsection{Scheduler patience.} With the default scheduler patience of 40 epochs, \texttt{GCN} attained markedly higher errors than \texttt{FNN} and \texttt{GCN\_khop} on the 4-input configuration (Tab.~\ref{tab:app_patience_4input}). Since \texttt{GCN} and \texttt{GCN\_khop} share the same architecture, and differ only in the graphs used during training and in the training batch size, this gap was investigated by varying the scheduler patience of \texttt{GCN}. The learning-rate scheduler (\texttt{ReduceLROnPlateau}) counts its patience in epochs, not in optimization steps. With batch size 2, on 21 full training meshes, \texttt{GCN} performs 11 updates per epoch, against several hundred for \texttt{GCN\_khop} on its sampled subgraphs; a patience of 40 epochs therefore corresponds to a very different number of updates for the two models. \texttt{GCN} was retrained with the patience increased from 40 to 400 epochs; the results obtained are those reported in Sec.~\ref{sec:results_rans}.

\begin{table}[htbp]
  \centering
  \caption{Effect of the scheduler patience on \texttt{GCN}, RANS
  framework, 4-input configuration (mixing length excluded), batch size fixed at 2. \texttt{FNN} and \texttt{GCN\_khop} included for reference.}
  \label{tab:app_patience_4input}
  \begin{tabular}{lccccc}
    \toprule
    Model & Epochs & Val.\ loss & err.\ training & err.\ validation & err.\ testing \\
    \midrule
    \texttt{FNN}                & 811  & $9.51\times10^{-6}$  & 0.1307 & 0.2223 & 0.2114 \\
    \texttt{GCN} (patience 40)  & 1102 & $2.053\times10^{-5}$ & 0.2400 & 0.3150 & 0.2766 \\
    \texttt{GCN} (patience 400) & 6000 & $8.079\times10^{-6}$ & 0.1051 & 0.1997 & 0.1809 \\
    \texttt{GCN\_khop}          & 408  & $7.206\times10^{-6}$ & 0.0927 & 0.1891 & 0.1886 \\
    \bottomrule
  \end{tabular}
\end{table}

Increasing the patience from 40 to 400 epochs brought \texttt{GCN} in line with the other two models. The error of \texttt{GCN} decreased on all three splits, by a factor between 1.5 and 2.3. With patience 400, \texttt{GCN} attains the lowest error of the three models on the testing set, and the second lowest on the training and validation sets, where it is 13\% and 6\% higher than \texttt{GCN\_khop} respectively; it is lower than \texttt{FNN} on all three splits. The validation loss follows the same behavior: at patience 40 it was more than twice as high as those of \texttt{FNN} and \texttt{GCN\_khop}, whereas at patience 400 it is equal to $8.079\times10^{-6}$, between \texttt{GCN\_khop}'s $7.206\times10^{-6}$ and \texttt{FNN}'s $9.51\times10^{-6}$. The differences among the three models at patience 400 are small, and are not sufficient to establish a ranking on the basis of a single training run. Note that \texttt{GCN} (patience 400) reached the maximum training epoch budget of 6000 without triggering early stopping which, at this patience, requires 800 epochs without improvement in the validation loss; so it is not established whether training had fully converged. The results above should be read as a lower bound on how much patience tuning can close, not necessarily the ceiling.

\section{Effect of the sampling depth $k$ on \texttt{GCN\_khop}, RANS dataset}\label{app:gcn_rans_khop_comparison}

As on the transport benchmark (Section~\ref{sec:sampling_speedup}), the effect of the sampling depth $k$ on \texttt{GCN\_khop} is assessed on the RANS dataset, comparing $k = 10$, $20$ and $50$ against the full-graph \texttt{GCN}. All three \texttt{GCN\_khop} configurations are trained at the default scheduler patience of 40 epochs, while \texttt{GCN} is trained at the increased patience of 400 epochs (Appendix~\ref{app:gcn_rans_4inputs_training}).

\begin{figure}[htbp]
    \centering
    \includegraphics[width=0.9\linewidth]{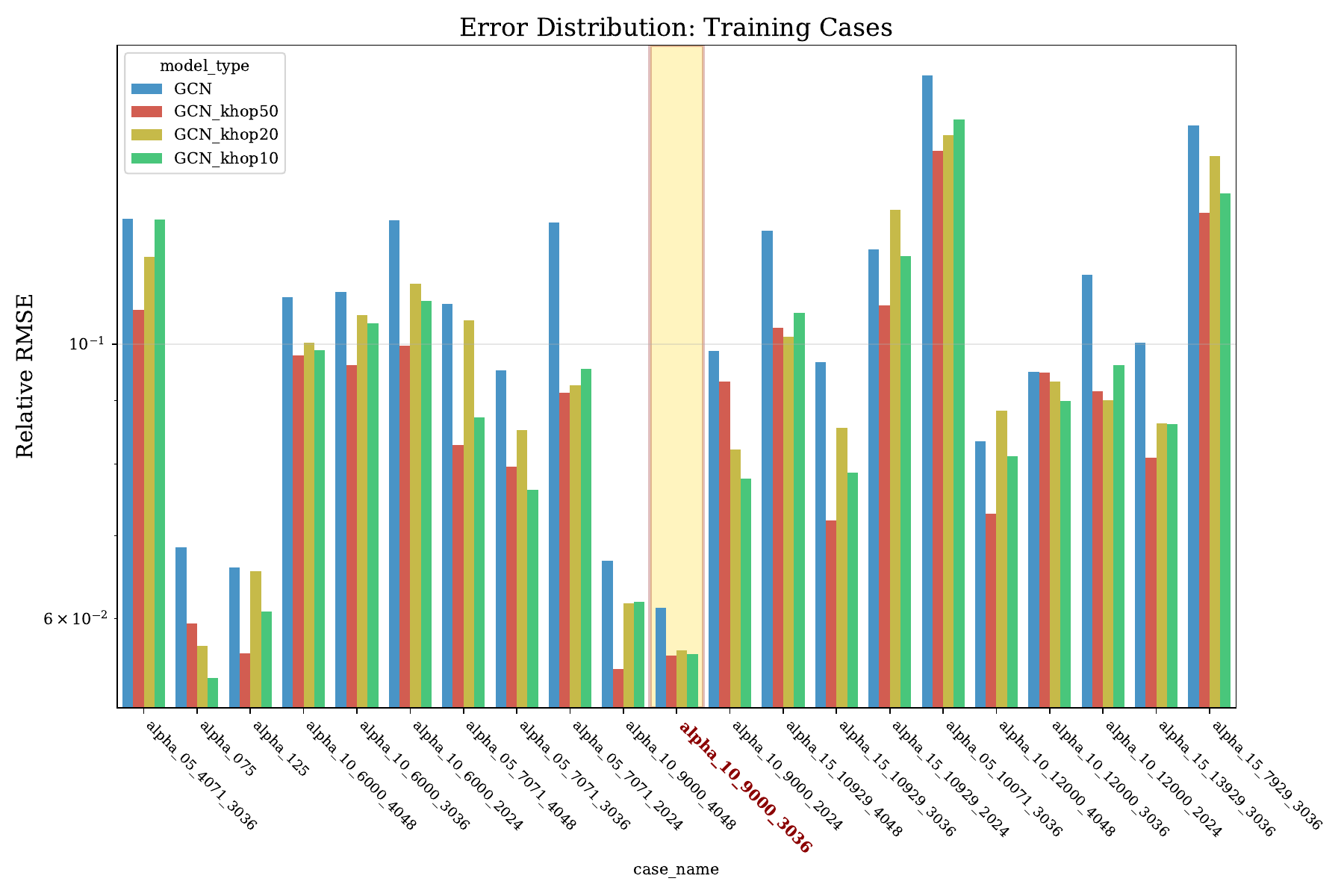}
    \caption{Relative RRMSE for \texttt{GCN} and \texttt{GCN\_khop} ($k = 10, 20, 50$) on every training configuration of the RANS dataset.}
    \label{fig:rans_khop_training_cases}
\end{figure}

\begin{figure}[htbp]
    \centering
    \includegraphics[width=0.9\linewidth]{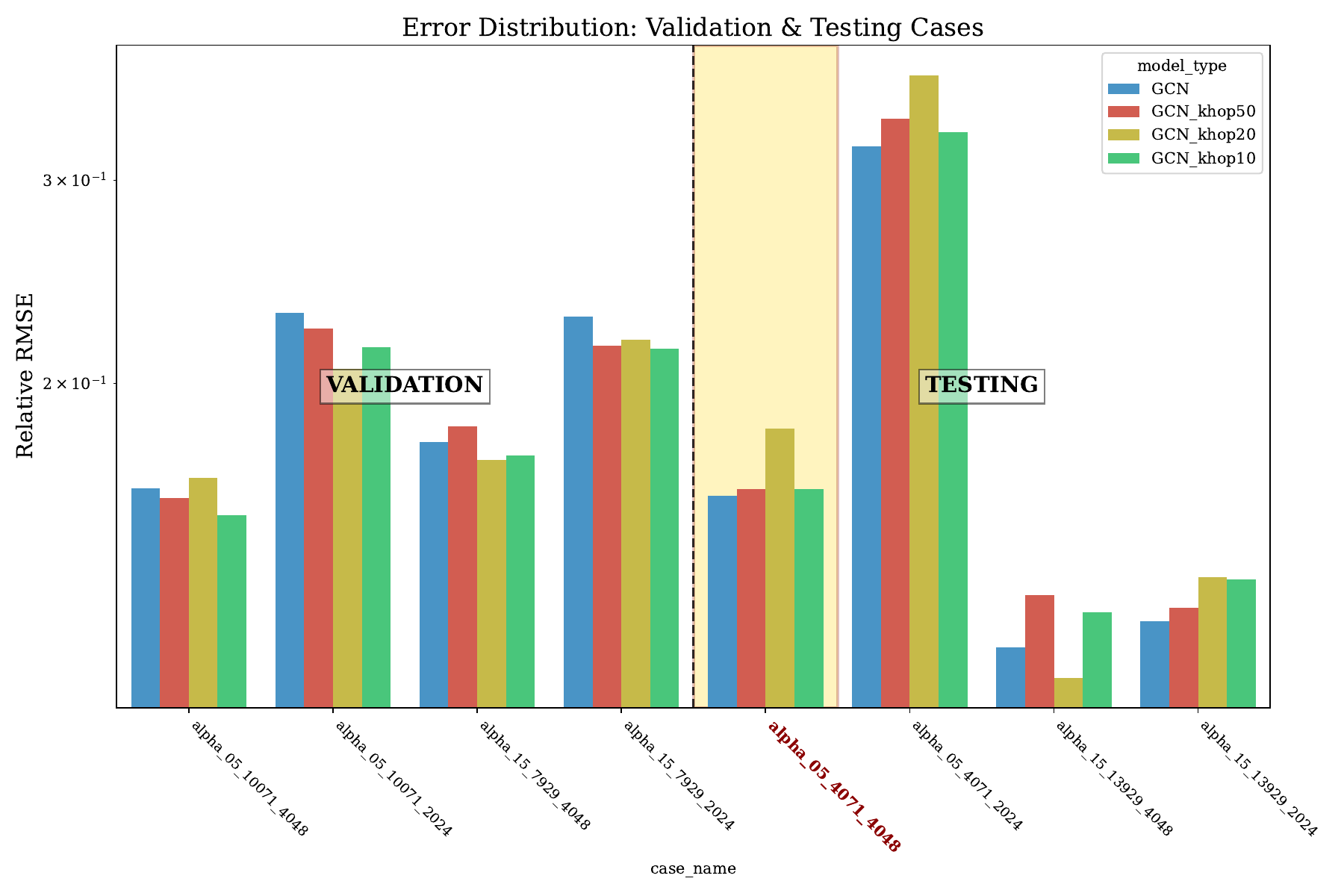}
    \caption{Relative RRMSE for \texttt{GCN} and \texttt{GCN\_khop} ($k = 10, 20, 50$) on every validation and testing configuration of the RANS dataset.}
    \label{fig:rans_khop_val_test_cases}
\end{figure}

Figure~\ref{fig:rans_khop_training_cases} shows the relative RRMSE for every training configuration. The four models attain errors of the same order on every configuration, and no model is consistently the most accurate: \texttt{GCN} is, on a majority of configurations, at or above the three sampled models, which in turn track each other closely. Figure~\ref{fig:rans_khop_val_test_cases} shows the same comparison on the validation and testing configurations. On the validation configurations the sampled models are on the whole slightly below \texttt{GCN}, while on the testing configurations \texttt{GCN} attains the lowest error on most of them, by small margins, with the lowest sampled error alternating among the three depths. The configurations highlighted in the figures are those shown in Figs.~\ref{fig:abs_err_nut_PH_train} and~\ref{fig:abs_err_nut_PH_test}.

Table~\ref{tab:rans_khop_mean_errors} reports the mean relative error over each split. On the training set, \texttt{GCN\_khop} with $k=50$ attains the lowest error, equal to 0.0889, followed by $k=10$ at 0.0927, $k=20$ at 0.0954 and \texttt{GCN} at 0.1051, which is 18\% higher than the best configuration. On the validation set, $k=10$ attains the lowest error, equal to 0.1891, followed by $k=20$ at 0.1905, $k=50$ at 0.1954 and \texttt{GCN} at 0.1997, 6\% higher than the best configuration. On the testing set, \texttt{GCN} attains the lowest error, equal to 0.1809, followed by $k=10$ at 0.1886, $k=50$ at 0.1901 and $k=20$ at 0.2000, which is 11\% higher than \texttt{GCN}. Among the sampled configurations, the errors differ by at most 7\% on any split, and their ranking changes from one split to another; $k=10$ is the best or the second best sampled configuration on all three splits.

\begin{table}[htbp]
    \centering
    \caption{Mean relative error of \texttt{GCN} and \texttt{GCN\_khop} ($k = 10, 20, 50$) on the RANS dataset, by split.}
    \label{tab:rans_khop_mean_errors}
    \begin{tabular}{lccc}
    \toprule
    \textbf{Model} & \textbf{Training} & \textbf{Validation} & \textbf{Testing} \\
    \midrule
    \texttt{GCN}          & 0.105136 & 0.199722 & 0.180936 \\
    \texttt{GCN\_khop10}  & 0.092687 & 0.189127 & 0.188570 \\
    \texttt{GCN\_khop20}  & 0.095389 & 0.190463 & 0.200005 \\
    \texttt{GCN\_khop50}  & 0.088920 & 0.195402 & 0.190074 \\
    \bottomrule
    \end{tabular}
\end{table}

Unlike on the transport benchmark, where the RRMSE on the test set decreased monotonically as $k$ decreased, on the RANS dataset the sampling depth has a limited effect on accuracy, and no consistent trend with $k$ emerges. Subgraph sampling yields errors of the same order as full-graph training at any of the three depths, within 18\% on every split. Since the differences among all configurations are small and each is obtained from a single training run, they should not be over-interpreted. The use of $k=10$ as the reference configuration in the main text is consistent with these results, as it is the best or the second best sampled configuration on every split.

\end{document}